\documentclass[%
 aip,
 amsmath,amssymb,
 reprint,%
]{revtex4-1}

\usepackage{float}
\usepackage{placeins}

\usepackage{graphicx}
\usepackage{caption}
\usepackage{subcaption}

\usepackage{dcolumn}% Align table columns on decimal point
\usepackage{bm}% bold math
\usepackage{mathtools} % for \vcentcolon and other math symbols
\usepackage[utf8]{inputenc}
\usepackage[T1]{fontenc}
\usepackage{mathptmx}
\usepackage{etoolbox}
\usepackage{amsthm}
\usepackage{xcolor}

\newtheoremstyle{break}% name
  {}%         Space above, empty = `usual value'
  {}%         Space below
  {\itshape}% Body font
  {}%         Indent amount (empty = no indent, \parindent = para indent)
  {\bfseries}% Thm head font
  {.}%        Punctuation after thm head
  {\newline}% Space after thm head: \newline = linebreak
  {}%         Thm head spec

\theoremstyle{break}
\newtheorem{definition}{Definition}[section]

\newcommand{\vbm}[1]{\boldsymbol{#1}}
\makeatletter
\def\@email#1#2{%
 \endgroup
 \patchcmd{\titleblock@produce}
  {\frontmatter@RRAPformat}
  {\frontmatter@RRAPformat{\produce@RRAP{*#1\href{mailto:#2}{#2}}}\frontmatter@RRAPformat}
  {}{}
}%
\makeatother
\begin{document}

\preprint{AIP/123-QED}

\title[Gyrofluid Magnetic Reconnection Beyond the $\delta F$ Approximation]{Gyrofluid Magnetic Reconnection Beyond the $\delta$F Approximation}
% Force line breaks with \\
\author{F. F. Locker} 
\affiliation{Department for Ion Physics and Applied Physics, University of Innsbruck, A-6020 Innsbruck, Austria%\\This line break forced with \textbackslash\textbackslash
}%

\author{M. Rinner}
\affiliation{Department for Ion Physics and Applied Physics, University of Innsbruck, A-6020 Innsbruck, Austria%\\This line break forced with \textbackslash\textbackslash
}

% Force line breaks with \\
\author{M. Held}
\affiliation{Department for Ion Physics and Applied Physics, University of Innsbruck, A-6020 Innsbruck, Austria%\\This line break forced with \textbackslash\textbackslash
}
\affiliation{Department of Mathematics and Statistics, UiT The Arctic University of Norway, 9037 Tromsø, Norway%\\This line break forced with \textbackslash\textbackslash
}%
\author{A. Kendl}
\affiliation{Department for Ion Physics and Applied Physics, University of Innsbruck, A-6020 Innsbruck, Austria%\\This line break forced with \textbackslash\textbackslash
}

\date{\today}% It is always \today, today,

\begin{abstract}
Magnetic reconnection in a two-dimensional system is studied employing a novel Full-$F$ gyrofluid model with arbitrary-wavelength polarization. In strongly magnetized plasmas, the long wavelength dimension parallel to the magnetic field can be separated from the much smaller perpendicular scales, motivating an isolated two-dimensional description. While previous studies have predominantly employed $\delta F$ gyrofluid models, the present Full-$F$ formulation is applied to simulate Harris-sheet magnetic reconnection with domain aspect ratios $L_{\hat{y}}/L_{\hat{x}}\leq16$, investigating tearing-mode growth, plasmoid formation, and the influence of finite ion Larmor radius (FLR) effects. In addition to a linear tearing-mode analysis, a non-modal stability analysis of the linearized system is performed. The evolution operator is shown to be strongly non-normal, exhibiting large condition numbers and extended pseudospectra that indicate the possibility of significant transient amplification, even in marginally stable regimes. Such transient amplification may facilitate the transition from linear tearing growth to rapid nonlinear acceleration. We focus on low-plasma-$\beta$ magnetic reconnection on the scale of the drift scale $\rho_s$ and incorporate ion FLR effects, placing the simulations in the context of magnetically confined fusion plasmas such as tokamaks. After discussing numerical resolution and convergence, we present a parameter scan varying the normalized electron skin depth and the ion-to-electron temperature ratio. Finally, the influence of aspect ratio, polarization model, and FLR effects on the reconnection dynamics and plasmoid formation is investigated.
\end{abstract}

\maketitle

\section{Introduction}

In magnetized plasmas, magnetic reconnection (MR) is known to play a crucial role in non-ideal magnetohydrodynamics (MHD), as magnetic field energy is converted into kinetic energy through the violation of flux conservation \citep{Boozer}. This process is fundamental to a wide range of systems, including stellar coronae, Earth’s magnetotail, and magnetically confined fusion devices \cite{Zweibel,Letsch}. Since the early studies by Sweet and Parker, and later by Petschek, research has been extended from the comparatively well understood resistive MR regime to its collisionless counterpart, with the non-resistive tearing instability being identified as a key moderator of the process \citep{Parker1,Parker2,Petschek,Uzdensky}. 

Early theoretical descriptions predicted a decline of the resistive tearing instability growth rate $\gamma_{\eta}\sim S^{-\frac{3}{5}}$ with increasing Lundquist number $S=Lv_A/\eta$, where $L$ denotes the characteristic system size, $v_A$ the Alfvén velocity, and $\eta = (\mu_0 \sigma)^{-1}$ the magnetic diffusivity, thereby limiting applicability to collisionless reconnection in the $\eta \rightarrow 0$ regime. It was later shown by Loureiro et al.\cite{Loureiro2} that Sweet–Parker current sheets become unstable to secondary tearing via the plasmoid instability at sufficiently large values of $S\sim 10^4$, leading to enhanced linear growth rates as well as nonlinear interactions, and rendering plasmoid-driven reconnection nearly independent of $S$ \cite{Baty,Uzdensky,Bhattacharjee,Comisso6,Comisso2}. Furthermore, plasmoid formation in such current sheets has been found to be independent of $\beta= \mu_0 p / B^2$, the ratio of plasma pressure $p$ to magnetic pressure $B^2/\mu_0$, in the high-$S$ regime \cite{Ziegler}. 

Current sheets thinner than an aspect ratio scaling as $a/L\sim S^{-1/3}$ (with $a,L$ denoting the sheet width and length) were proposed by Pucci et al.\cite{Pucci} to be inherently unstable, resulting in an ``ideal'' tearing mode characterized by reconnection rates independent of $S$ and explosive reconnection occurring on Alfvénic timescale. In low-$\beta  \ll 1$, high-$S$ plasmas with large aspect ratios $r_a=a/L$, island formation involving multiple X-points (or magnetic nulls in three dimensions) has been shown to increase reconnection rates between coalescing flux surfaces by enabling multiple simultaneous reconnection sites through the plasmoid instability \citep{Comisso2,Huang_2017,comisso2016visco}. Recently, it was highlighted by Morillo et al. \cite{Morillo} that plasmoid formation may be suppressed in numerical simulations if insufficient resolution is employed, even at high Lundquist numbers, underscoring the necessity of systematic convergence testing. This is particularly relevant for the present study, since arbitrary-wavelength polarization and $\rho_s$-scale structure require sufficient numerical resolution to distinguish physical plasmoid formation from numerical artifacts.

Gyrofluid and gyrokinetic models have previously been shown
to reproduce the essential collisionless mechanisms responsible
for fast magnetic reconnection, including electron inertia and
finite-Larmor-radius (FLR) effects
\cite{Ottavini,Grasso,Comisso,Biancalani,Scott3}.
In particular, gyrofluid studies demonstrated that fast
reconnection and nonlinear acceleration can occur even in
reduced fluid formulations while retaining key kinetic effects
\cite{Ottavini,Grasso,Comisso,Biancalani}.
Correspondingly, fast normalized reconnection rates
$E_{\rm rec}/(B_0 v_A)\sim 0.1$
have been reported in gyrofluid, gyrokinetic,
Hall-MHD, and fully kinetic simulations
\cite{Shay1999,Birn2001,Numata2011,Yamada2010,Granier3},
particularly for moderately low electron skin depth
$d_e=\sqrt{m_e/(\mu_0 e^2 n_e)}$
to drift scale
$\rho_s=\sqrt{T_e m_i}/(eB_0)$
ratios, $d_e/\rho_s \approx 1$.
However, many existing gyrofluid reconnection models employ
$\delta F$, Oberbeck--Boussinesq, or long-wavelength
approximations in the polarization equation. While these
approximations are often justified close to equilibrium and
for sufficiently large spatial scales, they may become
insufficient in strongly nonlinear reconnection regimes with
finite-wavelength polarization effects and large density
fluctuations.

As a motivation for the present study, it is worth recalling
that in toroidal fusion plasmas such as tokamaks, the plasma is
typically characterized by high Lundquist numbers, and
resonant flux surfaces are susceptible to tearing
instabilities and subsequent magnetic island formation
\cite{Boozer2}. These processes are generally associated with a
degradation of plasma confinement through the flattening of
radial pressure profiles and enhanced cross-field transport.
Although the present work is restricted to a two-dimensional
slab geometry, understanding the underlying collisionless
reconnection physics provides an important step toward more
realistic fusion-relevant configurations.

In the present work, the recently published code GREENY
(Gyrofluid Reconnection with Extended Electromagnetic
Nonlinearity) \cite{Locker3} is employed. GREENY is a
two-dimensional Full-$F$ gyrofluid simulation framework
designed for the investigation of collisionless magnetic
reconnection in parameter regimes relevant to fusion plasmas.
A two-dimensional geometry is often employed as a first
approximation for tokamak plasmas, although more detailed
modelling requires the inclusion of toroidal curvature,
trapped-particle effects, and global equilibrium gradients.
Here, the term ``collisionless'' refers to the neglect of
parallel resistive (e.g.\ Spitzer) terms, such that
$\eta_S \rightarrow 0$, and reconnection is instead mediated
by non-ideal effects such as electron inertia and weak numerical
or hyper-viscosity. In contrast to many reduced gyrofluid
approaches, arbitrary-wavelength polarization is retained and
total densities are evolved self-consistently in time.

\begin{table}[h]
    \centering
    \begin{tabular}{|l|l|}
    \hline
      species temperature  & $ T_z$ \\
       \hline
       species mass  & $m_z$\\
              \hline
        charge number  & $Z_z$\\
                      \hline
        perpendicular gradient length  & $L_\perp$\\
       \hline
           charge  & $q_z:=eZ_z$\\
       \hline
               drift scale  & $\rho_{s}:=\sqrt{T_e m_i}/(eB_0)$\\
       \hline
               electron skin depth  & $d_{e}:=\sqrt{m_e/(\mu_0 e^2 n_e)}$\\
       \hline
       temperature ratio  & $\tau_z:=\frac{T_z}{Z_z T_e}$\\
       \hline
       mass ratio & $\mu_z:=\frac{m_z}{Z_z m_i}$\\
       \hline
         electron beta  & $\beta=\mu_0 N_{e0} T_e/B_0^2$\\
                \hline
       sound speed  & $c_{s}:=\sqrt{T_e/m_i}$\\
          \hline
        $E \times B$ velocity & $|\vbm{U_{E}}|=1/B_0| \nabla_{\perp} \hat{\psi}_z|$\\
                   \hline
         perpendicular Poisson bracket & 
         $\begin{aligned} 
         1/B_0 (\vbm{\hat{b}} \times \nabla_{\perp} f ) \cdot \nabla g 
         = 1/B_0 [f,g]_{\perp}
         \end{aligned}$\\
       \hline
       \end{tabular}
    \caption{Physical quantities, system parameters, and identities.}
        \label{tab:param} 
\end{table}

This formulation defines a Full-$F$, non-Oberbeck--Boussinesq
(NOB) gyrofluid model, governed by the system parameters
listed in Table~\ref{tab:param}, with a constant species
gyro-radius
$\rho_z = \sqrt{T_z m_z}/(eB)=\mathrm{const.}$
and a locally constant guide field $B=B_0$ \cite{Kendl}.
Consequently, the model is isothermal and, while densities are
evolved as total quantities, the magnetic-field fluctuations remain perturbative in amplitude \cite{Held}. A reduced
Oberbeck--Boussinesq ($\delta F$) limit is also available, in
which densities $N_z$ are decomposed into a constant background
$N_0$ and perturbations $\delta N_z$, together with an
OB-approximated polarization equation. Perpendicular magnetic
field fluctuations are computed self-consistently through
Ampère's law. The numerical implementation and associated
modeling assumptions are discussed in detail in Ref.~\citenum{Locker3}.

While typically $\delta F$ gyrofluid formulations are obtained using perturbative gyrokinetic expansions around a prescribed equilibrium, Full-$F$ models are derived from moment equations of the Full-$F$ gyrokinetic Vlasov equation
\cite{Madsen,ScottV2}. The primary difficulty lies in the
treatment of quasi-neutrality and Ampère's law, which require
additional approximations in order to obtain a closed system of
equations. These approximations become particularly relevant in
strongly nonlinear reconnection scenarios, where large density
fluctuations, finite-wavelength effects, and nonlinear
polarization dynamics may invalidate assumptions commonly used
in reduced models.

Within the scope of this work, collisionless magnetic
reconnection is investigated in the described Full-$F$
gyrofluid framework, with a focus on a low-$\beta$ parameter
regime and perpendicular scales comparable to the drift scale,
$L_\perp \sim \rho_s$. This regime is chosen in order to
emphasize finite-wavelength polarization and ion-FLR effects,
for which long-wavelength and Oberbeck-Boussinesq
approximations may become insufficient.

Following the presentation of the model equations in
Section~\ref{model}, a linear perturbation analysis is performed
in Section~\ref{LinAn} in order to derive an analytical estimate
for the linear tearing growth rate
$\hat{\gamma}_{\mathrm{lin}}$.
Subsequently, the conditioning and transient behavior of the
linearized system are investigated by means of a non-modal
analysis in Section~\ref{nomo}.
Section~\ref{plamoresu} then presents nonlinear simulations
of Harris-sheet magnetic reconnection under different
polarization assumptions, including comparisons between
long-wavelength and arbitrary-wavelength polarization models,
as well as between $\delta F$ and Full-$F$ formulations.
The influence of aspect ratio, electron skin depth
$\hat d_e = d_e/\rho_s = \sqrt{\mu_e}/\sqrt{\beta}$,
and ion temperature effects on reconnection dynamics and
plasmoid formation is investigated systematically.
Finally, conclusions and a summary of the results are presented
in Section~\ref{Concl}.

\section{Model Equations}\label{model}
\subsection{Full-$F$ collisionless MR model}
The collisionless magnetic reconnection model, introduced in Ref.~\citenum{Locker3}, is derived from the Full-$F$ gyrofluid formulation of Madsen~\cite{Madsen} together with Held et al.~\citenum{Held2}. For the sake of completeness, the model and its limits are presented again. The low-$\beta$ limit, $\beta \ll 1$, of the Full-$F$ model is considered, all curvature effects are neglected, and a uniform guide magnetic field $\vbm{B} = B_0 \vbm{\hat{e}}_{\hat{\parallel}}$ is assumed. The evolution equations for the density and canonical momentum are 
\begin{equation}
\frac{\partial}{\partial t}N_z +\frac{1}{B}\{\psi_z,N_z\}_\perp -N_z\frac{1}{B}\{\xi_{1,z},U_z\}_\perp- U\frac{1}{B}\{\xi_{1,z},N_z\}_\perp = \Lambda
\end{equation}
\begin{align}
&\frac{m_zN_z}{q_z}\frac{\partial}{\partial t}U_z + N_z \frac{\partial}{\partial t} \xi_{1,z}= -\frac{m_zN_z}{q_zB}\{\psi_z,U_z\}_\perp +\notag\\
 &+N_z\frac{1}{B}\{\xi_{1,z},\psi_z\}_\perp +\frac{m_zU_zN_z}{q_zB}\{\xi_{1,z},U_z\}_\perp + \frac{T_z}{q_zB}\{\xi_{1,z},N_z\}_\perp +\frac{1}{q_z}\Lambda
\end{align}

for each species, where $\psi_{1,z} :=\Gamma_{1,z} \phi$ and $\xi_{1,z} :=\Gamma_{1,z} A_{\parallel}$ and $\Lambda$ denotes sources and sinks (e.g. subgrid-dissipation). These evolution equations are normalized (denoted by $\hat{\cdot}$) according to the Bohm normalization scheme shown in Table \ref{tab:norm}.

\begin{table}[htp]
    \centering
    \begin{tabular}{|l|l|}
    \hline
       time  & $ t:=\frac{\hat{L}_{\hat{\perp}}}{c_s}\hat{t}$\\
       \hline
       gyrocenter number density  & $N_z:=N_0 \hat{N}_z$\\
       \hline
       Poisson bracket  & $ \{\cdot, \cdot \}_{\perp}:=\rho_s^{-2}[\cdot,\cdot]_{\hat{\perp}}$\\
       \hline
        magnetic-field magnitude  & $B:=B_0 \hat{B}$\\
       \hline
         generalized electric potential  & $\psi_z:=\frac{T_e}{e}\hat{\psi}$\\
       \hline
       parallel velocity  & $U_z:=c_s \hat{U}_z$\\
       \hline
        parallel vector potential & $A_{\parallel}:=( B_0 \rho_{s}) \hat{\tilde{A}}_{\hat{\parallel}}$\\
       \hline
        perpendicular gradient operator  & $\nabla_\perp:=\rho_s^{-1}\hat{\nabla}_{\perp}$\\
       \hline
       \end{tabular}
    \caption{Bohm normalization scheme of the dependent fields.}
    \label{tab:norm} 
\end{table}

The corresponding system constants and parameters are listed in Table~\ref{tab:param}. We decompose the magnetic-field unit vector as $\vbm{\bar{b}}=\vbm{b}_0+\vbm{\tilde{b}}$, where the fluctuating component is given by $\vbm{\tilde{b}}=\frac{1}{B_0}\nabla A_{{\parallel}}\times\vbm{\hat{b}}$ and the static guide field is $\vbm{b}_0$. The normalized parallel gradient of a field $\hat{f}$ becomes
\begin{align}\label{gradient}
\hat{\bar{\nabla}}_{\hat{\parallel}} \hat{f}
:=\hat{\nabla}_{\hat{\parallel}} \hat{f}
+\vbm{\tilde{b}}\cdot \hat{\nabla} \hat{f}
= \hat{\nabla}_{\hat{\parallel}} \hat{f}
-\frac{\beta}{ \hat{B \delta}}[\hat{\tilde{A}}_{\hat{\parallel}},\hat{f}]_{\hat{\perp}}
\end{align}
where the second term is magnetic flutter. In this normalization, the parameter $\delta = \rho_s/L_{\perp}$ scales the Poisson brackets and hence the system dynamics. Since the characteristic gradient length is of order $\rho_s$, the dimensionless scaling parameter is set to $\delta=1$. The gyrocenter $E\times B$ drift is determined by the gyroaveraged and polarization components of the gyrofluid potential $\hat{\psi}_z=\hat{\psi}_{1,z}+\hat{\psi}_{2,z}$ \cite{Held3},
\begin{align}
\hat{\psi}_{1,z} :=\hat{\Gamma}_{1,z} \hat{\phi}, \\
\hat{\psi}_{2,z}:= - \frac{\mu_z}{2} \left|\hat{\nabla}_{\perp}\sqrt{\hat{\Gamma}_{0,z}} \hat{\phi} \right|^2.
\end{align}
The exact form of the operators depends on the model assumptions summarized in Table~\ref{polapproxy}. For the remainder of this work, $\beta$ is absorbed into $\hat{A}_{{\hat{\parallel}}}=\beta \hat{\tilde{A}}_{\hat{\parallel}}$. The Padé-approximated gyroaverage and screening operators are \citep{Held2}
\begin{align}
\hat{\Gamma}_{0,i}&:=\frac{1}{1-\tau_i \mu_i \hat{\Delta}_{\perp}},&
\hat{\Gamma}_{1,i}&:= \frac{1}{1-\frac{\tau_i}{2} \mu_i\hat{\Delta}_{\perp}}
.
% ,\\
% \hat{\Gamma}_{0,e}&:=\frac{1}{1+\mu_e \tau_e \hat{\Delta}_{\perp}},&
% \hat{\Gamma}_{1,e}&:= \frac{1}{1+\frac{\mu_e}{2} \tau_e \hat{\Delta}_{\perp}}.
\end{align}
Since $|\mu_e|\ll 1$ for proton–electron plasmas, electron FLR effects may be neglected, which implies $\hat{\psi}_e=\hat{\phi}$, \(\hat{\Gamma}_{0,e}=\hat{\Gamma}_{1,e}=1\), $\hat{N}_e = \hat{n}_e$ and $\hat{U}_e = \hat{u}_e$. The normalized and implemented Full-$F$ equations are
\begin{align}\label{FFmodel}
    \frac{\partial}{\partial \hat{t}} \hat{n}_e &=
    -\left[\hat{\phi},\,\hat{n}_e\right]_{\hat{\perp}} 
    + \left[\hat{A}_{\hat{\parallel}},\,\hat{n}_e\,\hat{u}_e\right]_{\hat{\perp}},
    \\
    \frac{\partial}{\partial \hat{t}} \hat{N}_i &=
    -\left[\hat{\psi}_i,\,\hat{N}_i\right]_{\hat{\perp}} 
    + \left[\hat{\Gamma}_1 \hat{A}_{\hat{\parallel}},\,\hat{N}_i\,\hat{U}_i\right]_{\hat{\perp}},
    \\ \nonumber
    \frac{\partial}{\partial \hat{t}}\left(\hat{u}_e+\frac{1}{\mu_e}\,\hat{A}_{\hat{\parallel}}\right) =
    &-\left[\hat{\phi},\,\hat{u}_e
     +\frac{1}{\mu_e}\,\hat{A}_{\hat{\parallel}}\right]_{\hat{\perp}}+
    \\ &+\left[\hat{A}_{\hat{\parallel}},\,\frac{1}{2}\,\hat{u}_e^2\right]_{\hat{\perp}}
    -\frac{1}{\mu_e}\left[\hat{A}_{\hat{\parallel}},\,\ln(\hat{n}_e)\right]_{\hat{\perp}},
    \\ \nonumber
    \frac{\partial}{\partial \hat{t}}\left(\hat{U}_i+\hat{\Gamma}_1 \hat{A}_{\hat{\parallel}}\right) =
    &-\left[\hat{\psi}_i,\,\hat{U}_i +\hat{\Gamma}_1 \hat{A}_{\hat{\parallel}}\right]_{\hat{\perp}}
    +\\&+ \left[\hat{\Gamma}_1 \hat{A}_{\hat{\parallel}},\,\frac{1}{2}\,\hat{U}_i^2\right]_{\hat{\perp}}
    + \tau_i\left[\hat{\Gamma}_1 \hat{A}_{\hat{\parallel}},\,\ln(\hat{N}_i)\right]_{\hat{\perp}}.
\end{align}

With the $\mathbf E\times\mathbf B$ drift and the perpendicular magnetic flutter~\ref{gradient}, the species advection velocities are
\begin{align}
\hat{\vbm{V}}_e := \hat{\vbm{u}}_E + \hat{u}_e\,\vbm{\tilde{b}},\\
\hat{\vbm{V}}_i := \hat{\vbm{u}}_E + \hat{U}_i\,\vbm{\tilde{b}}.
\end{align}

where $\nabla \cdot \hat{V}_z = 0$. Hence, the equations in advective form are

\begin{align}
 \frac{\partial}{\partial \hat{t}} \hat{n}_e
 &= - \hat{\mathbf{V}}_e \cdot \nabla \hat{n}_e
 \;-\;
 \hat{n}_e\,\hat{\boldsymbol{b}}\cdot\nabla \hat{u}_e,
\label{eq:ne_adv}
\\[0.5em]
 \frac{\partial}{\partial \hat{t}} \hat{N}_i
 &= - \hat{\mathbf{V}}_i \cdot \nabla \hat{N}_i
 \;-\;
 \hat{N}_i\,\hat{\boldsymbol{b}}\cdot\nabla \hat{U}_i,
\label{eq:Ni_adv}
\\[0.5em]
 \frac{\partial}{\partial \hat{t}}
 \left(
   \hat{u}_e+\frac{1}{\mu_e}\,\hat{A}_{\parallel}
 \right)
&= - \hat{\mathbf{V}}_e\cdot\nabla \hat{u}_e
\;+\;
\frac{1}{\mu_e}\,
\hat{\boldsymbol{b}}\cdot\nabla
\left(
  \ln \hat{n}_e - \hat{\phi}
\right),
\label{eq:ue_adv}
\\[0.5em]
 \frac{\partial}{\partial \hat{t}}
 \left(
   \hat{U}_i+\hat{\Gamma}_1 \hat{A}_{\parallel}
 \right)
&= - \hat{\mathbf{V}}_i\cdot\nabla \hat{U}_i
\;-\;
\tau_i\,\hat{\boldsymbol{b}}\cdot\nabla\ln \hat{N}_i
\;-\;
\hat{\boldsymbol{b}}\cdot\nabla \hat{\psi}.
\label{eq:Ui_adv}
\end{align}
Here we used $\hat{N}_e \approx \hat{n}_e$ and $\hat{U}_e \approx \hat{u}_e$, valid for electron mass ratios $|\mu_e|\ll 1$ at $\rho_s$ scales and small $\beta$. This regime is characteristic of fusion plasma simulations dominated by drift-wave dynamics interacting with reconnecting magnetic field structures. While in a turbulent setting, the implementation of equations \ref{FFmodel} requires a subgrid dissipation \cite{smith1997eddy,scott1992mechanism,camargo2000nonmodal} one generally also needs regularization for centered differencing. This is done by transforming
\begin{align}
    \frac{\partial}{\partial \hat{t}}\rightarrow \frac{\partial}{\partial \hat{t}}+\nu \hat{\nabla}_{\hat{\perp}}^4.
\end{align}
For the problem of magnetic reconnection one can either take the numerical diffusion into account \cite{Biancalani} or choose the scalar parameter $\nu$ small enough to prevent a strong influence on the reconnecting current sheet \cite{Locker3}. In the presented setting, a small hyperviscosity parameter of $\nu \propto 10^{-6}, 10^{-7}$ is used, further discussed in \ref{tearing}. 

The system is closed by Ampère’s law,
\begin{align}
    -\frac{1}{\beta}\,\hat{\Delta}_{\perp} \hat{A}_{\hat{\parallel}} 
    = -\hat{n}_e\,\hat{u}_e + \hat{\Gamma}_1(\hat{N}_i\,\hat{U}_i),
\end{align}
and the arbitrary-wavelength, non-Oberbeck–Boussinesq (NOB) polarization equation,
\begin{align}
\hat{\vbm{\nabla}} \cdot \left[\sqrt{\hat{\Gamma}_0}\,
\frac{\hat{N}_i}{\hat{B}^2}\sqrt{\hat{\Gamma}_0}\,
\hat{\nabla}_{\hat{\perp} }\hat{\phi}\right]
= -\sum_z Z_z \hat{\Gamma}_{1,z} \hat{N}_z.
\end{align}

The polarization equation may be written as
\begin{align}
    \sum_z Z_z\hat{N}_z - \hat{\vbm{\nabla}} \cdot \hat{\mathbf{P}} = 0,
\end{align}
where $\hat{\mathbf{P}}=\hat{\mathbf{P}}_1+\hat{\mathbf{P}}_2$ \citep{Held3,Kendl} and
\begin{align}
    \hat{\mathbf{P}}_1 &:= -\sum_z Z_z \hat{\nabla}_{\perp}\hat{\Gamma}_{1,z}\left(\mu_z \tau_z \hat{N}_z/2\right),\\
    \hat{\mathbf{P}}_2 &:= -\sum_z \left(\sqrt{\hat{\Gamma}_{0,z}}\, Z_z\mu_z \frac{\hat{N}_z}{\hat{B}^2}\sqrt{\hat{\Gamma}_{0,z}}\, \hat{\nabla}_{\hat{\perp} }\hat{\phi}\right).
\end{align}
 Here the second-order Padé approximation $\hat{\Gamma}_1\approx\sqrt{\hat{\Gamma}_0}$ is used\cite{Held2}, together with 
 \begin{align}
 -\hat{\vbm{\nabla}} \cdot \mathbf{P}_1 = \sum_z Z_z (\hat{\Gamma}_1 - 1)\hat{N}_z.
 \end{align}
 The respective OB and long wavelength (LWL) approximation of the polarization equation are summed up in table \ref{polapproxy}. The normalized total energy is
\begin{equation}
\begin{split}
\hat{E}_{\mathrm{tot}} = 
\sum_z \int_V dV \Big(&
Z_z \tau_z \hat{N}_z \ln(\hat{N}_z)
- Z_z  \hat{N}_z \hat{\psi}_{2,z} \\
&+ Z_z \mu_z \tfrac{1}{2}\hat{N}_z \hat{U}_z^2
+ \tfrac{|\hat{\nabla}_\perp\hat{A}_{\hat{\parallel}}|^2}{2\beta}
\Big)
\end{split}
\end{equation}
which consists of the Helmholtz free energy $\hat{E}_n$, the $E\times B$ energy $\hat{E}_{E\times B}$, the parallel energy $\hat{E}_{{\hat{\parallel}}}$, and the perturbed magnetic energy $\hat{E}_m$. The chosen normalization targets turbulence at the scale of $\rho_s$, thereby defining the resolvable range of fluctuations and implying that only ion FLR effects are retained. These effects become significant mainly in large–aspect ratio systems and in turbulent magnetic-reconnection scenarios \cite{Biancalani}.

\subsection{Oberbeck-Boussinesq approximated equations}
One can recover the OB (\(\delta F\)) limit, assuming a constant background density $N_0$ and small perturbations $\hat{\tilde{N}}_z$, substituting $\hat{N}_z=1 + \hat{\tilde{N}}_z$  in the Full-$F$ equations \cite{ScottV2}. Further simplification is obtained by assuming $\hat{\tilde{n}}_z \ll \hat{N}_0$ (or, in normalized form, $\hat{\tilde{n}}_z \ll 1$), neglecting triple non-linearities and employing the $\delta F$ polarization equation.

\begin{align}
    %1] \hspace{2cm}
    \frac{\partial}{\partial \hat{t}} \hat{\tilde{n}}_e &=
    -\left[\hat{\phi},\,\hat{\tilde{n}}_e\right]_{\hat{\perp}}
    + \left[\hat{A}_{\hat{\parallel}},\,\,\hat{u}_e\right]_{\hat{\perp}},
    \label{eq:df1}
    \\
    %2] \hspace{2cm}
    \frac{\partial}{\partial \hat{t}}  \hat{\tilde{N}}_i &= 
    -\left[\hat{\psi},\,\delta \hat{\tilde{N}}_i\right]_{\hat{\perp}} +
    \left[\hat{\Gamma}_1 \hat{A}_{\hat{\parallel}},\,\,\hat{U}_i\right]_{\hat{\perp}},    \label{eq:df2}
    \\
    %3] \hspace{2cm}
    \frac{\partial}{\partial \hat{t}}\left(\hat{u}_e+\frac{1}{\mu_e}\,\hat{A}_{\hat{\parallel}}\right) &=
    -\left[\hat{\phi},\,\hat{u}_e+\frac{1}{\mu_e}\,\hat{A}_{\hat{\parallel}}\right]_{\hat{\perp}} - \nonumber \\ 
   & -\frac{1}{\mu_e}\,\left[\hat{A}_{\hat{\parallel}},\,\hat{\tilde{n}}_e\right]_{\hat{\perp}},
  \label{eq:df3}
    \\
    %4] \hspace{2cm}
    \frac{\partial}{\partial \hat{t}}\left(\hat{U}_i+\hat{\Gamma}_1 \hat{A}_{\hat{\parallel}}\right) &=
    -\left[\hat{\psi},\,\hat{U}_i+\hat{\Gamma}_1 \hat{A}_{\hat{\parallel}}\right]_{\hat{\perp}} + \nonumber \\
   +& \tau_i\,\left[\hat{\Gamma}_1 \hat{A}_{\hat{\parallel}},\,\hat{\tilde{N}}_i\right]_{\hat{\perp}},
   \label{eq:df4}
    \\
    %5] \hspace{2cm}
    -\frac{1}{\beta}\,\hat{\Delta}_{\perp} \hat{A}_{\hat{\parallel}} &=
    -\,\hat{u}_e + \hat{\Gamma}_1\left(\,\hat{U}_i\right),
    \label{eq:df5}
    \\
    %6] \hspace{2cm}
 \frac{1}{\tau_i}(\hat{\Gamma}_0-1)\hat{\phi} &=  \hat{\tilde{n}}_e - \hat{\Gamma}_1 \hat{\tilde{N}}_i
 \label{eq:df6}
\end{align}
 Note, that the Full-$F$ and Full-$F$ + OB polarization equations coincide only in their respective long-wavelength limit (LWL) (Table \ref{polapproxy}). These equations match the model derived by Scott \cite{ScottV2}. A key difference between the Full-$F$ and Full-$F$ + OB systems is the presence of cubic nonlinearities in all evolution equations.
 
 \begin{table}   
    \centering
    \begin{tabular}{|l|l|l|}
    \hline 
    \textbf{GF model}& $\hat{\textbf{P}}_2$ & $\hat{\psi}_2$\\
    \hline 
    Full-$F$ & $-\sum_z\sqrt{\hat{\Gamma}_0} \mu_z Z_z \frac{\hat{N}_z}{\hat{B}_0^2} \sqrt{\hat{\Gamma}_0} \hat{\nabla}_{\perp} \hat{\phi} $& $-\mu_z |\hat{\nabla}_{\perp} \sqrt{\hat{\Gamma}}_0 \hat{\phi} |^2$\\
    \hline 
    Full-$F$ + OB & $ -\sum_z \mu_z Z_z\frac{\hat{N}_0}{\hat{B}_0^2} \hat{\Gamma}_0 \hat{\nabla}_{\hat{\perp} }\hat{\phi}$& 0\\
    \hline 
        Full-$F$ + LWL & $ - \sum_z \mu_z Z_z \frac{\hat{N}_z}{\hat{B}_0^2} \hat{\nabla}_{\hat{\perp} }\hat{\phi} $& $-\mu_z |\hat{\nabla}_{\hat{\perp} }  \hat{\phi} |^2$\\
        \hline 
  Full-$F$ + OB + LWL & $- \sum_z \mu_z Z_z \frac{\hat{N}_0}{ \hat{B}_0^2} \hat{\nabla}_{\hat{\perp} }\hat{\phi}$& 0\\
    \hline
    \end{tabular}
    \caption{Long wavelength (LWL) and Oberbeck-Boussinesq (OB) approximation of the polarization density  $\hat{\textbf{P}}_2$ and polarization part of the gyrofluid potential $\hat{\psi}_2$ for different models.}
     \label{polapproxy}    
\end{table}

\section{Stability Analysis}
Section~\ref{LinAn} and Section~\ref{nomo} concern the linearized system. 
The normal-mode analysis describes the asymptotic modal growth rate, whereas the non-modal analysis characterizes possible finite-time amplification caused by the non-normality of the linearized evolution operator. 
These mechanisms are distinct from the nonlinear explosive reconnection phase discussed in Section~\ref{tearing}. 
However, in the Full-$F$ system the nonlinear evolution continuously modifies the effective linearized operator around the instantaneous state of the plasma. 
Consequently, transient amplification associated with non-normality may persist beyond the initial linear phase and contribute to the onset of nonlinear explosive reconnection.

\subsection{Linearized System of Equations}\label{LinAn}
To obtain an analytical estimate $\hat{\gamma}_{\mathrm{lin}}$ for the tearing instability growth rate $\hat{\gamma}$, the approach of Tassi et al.~\cite{Tassi} is followed. The Full-$F$ equations are linearized in the cold-ion limit $\tau_i = T_i/T_e \approx 0$, whereby the gyroaveraging operators $\hat{\Gamma}_1$ and $\hat{\Gamma}_0$ are eliminated (yielding, e.g. $\hat{\psi} = \hat{\phi}$). A harmonic ansatz of the form
\begin{align}
\hat{F}(x,y,t) = \hat{F}_{\mathrm{eq}}(\hat{x}) + \tilde{F}(\hat{x})\,\exp\!\bigl(\mathrm{i}(\hat{k}\hat{y} - \hat{\omega} \hat{t})\bigr),
\end{align}
with $\hat{\omega} = \mathrm{i}\hat{\gamma}$ and $\hat{F}$ representing one of the six fields, is employed. The analysis is restricted to the small-growth-rate regime $\hat{\gamma} \ll 1$, using the following equilibrium functions, which are characteristic of a Harris-sheet initial condition.

\begin{align}
    \hat{n}_{e,\mathrm{eq}}(\hat{x}) &= N_0 = \mathrm{const.},
    \label{eq:4.2_init_ne}
    \\
    \hat{N}_{i,\mathrm{eq}}(\hat{x}) &= N_0,
    \label{eq:4.2_init_Ni}
    \\
    \hat{U}_{e,\mathrm{eq}}(\hat{x}) &= \frac{1}{\beta N_0}\,\nabla^2_\perp \hat{A}_{\hat{\parallel},\mathrm{eq}}(\hat{x}),
    \label{eq:4.2_init_Ue}
    \\
    \hat{U}_{i,\mathrm{eq}}(\hat{x}) &= 0,
    \label{eq:4.2_init_Ui}
    \\
    \hat{\phi}_\mathrm{eq}(\hat{x}) &= 0,
    \label{eq:4.2_init_phi}
    \\
    \hat{A}_{\hat{\parallel},\mathrm{eq}}(\hat{x}) &= \beta A_0\,a_{\hat{\parallel},\mathrm{eq}}(\hat{x}) =-\beta A_0\,\ln(\cosh(\hat{x})),
    \label{eq:4.2_init_A}
\end{align}

Analogous to Refs.~\cite{Tassi,Furth}, an analytical estimate
for the tearing stability parameter $\hat{\Delta}'$ is obtained
in the marginally stable limit
$\hat{\Delta}' \ll 1$
(Appendix~\ref{AppendixA}) by matching the inner and outer
solutions,
\begin{align}\label{dispersion_relation}
\begin{split}
\hat{\Delta}'_{\mathrm{outer}}
&=
2\left(
\frac{1}{\hat{k}_{\hat y}}
-
\hat{k}_{\hat y}
\right)
=
\hat{\Delta}'_{\mathrm{inner}}
\\
&=
\beta N_0
\sqrt{
\frac{\hat{\gamma}_{\mathrm{lin}}}
{A_0 \hat{k}_{\hat y}}
}
\Bigg[
\frac{2.12}{\sqrt{\beta}}
+
\pi
\left(
\frac{
\varphi^{-1}
+
\mu_e N_0^2
}
{
-\mu_e N_0^2
}
\right)
\left(
\frac{
\hat{\gamma}_{\mathrm{lin}}
}
{\iota}
\frac{
-\mu_e N_0^2
}
{
A_0 \hat{k}_{\hat y}
}
\right)^{1/2}
\Bigg].
\end{split}
\end{align}
Here,
\begin{align}
\iota &= \beta^2(A_0^2\mu_e + N_0^2),
\\
\varphi^{-1}
&=
N_0^2
\left(
1
-
\mu_e
\left(
1+\frac{2}{\beta N_0}
\right)
\right),
\end{align}
and
$\hat{k}_{\hat y}$
denotes the normalized wave number in the $\hat y$ direction.

\begin{center}
\begin{figure}[htp]
\centering
  \includegraphics[trim=0 0 0 0,clip,width=0.45\textwidth]{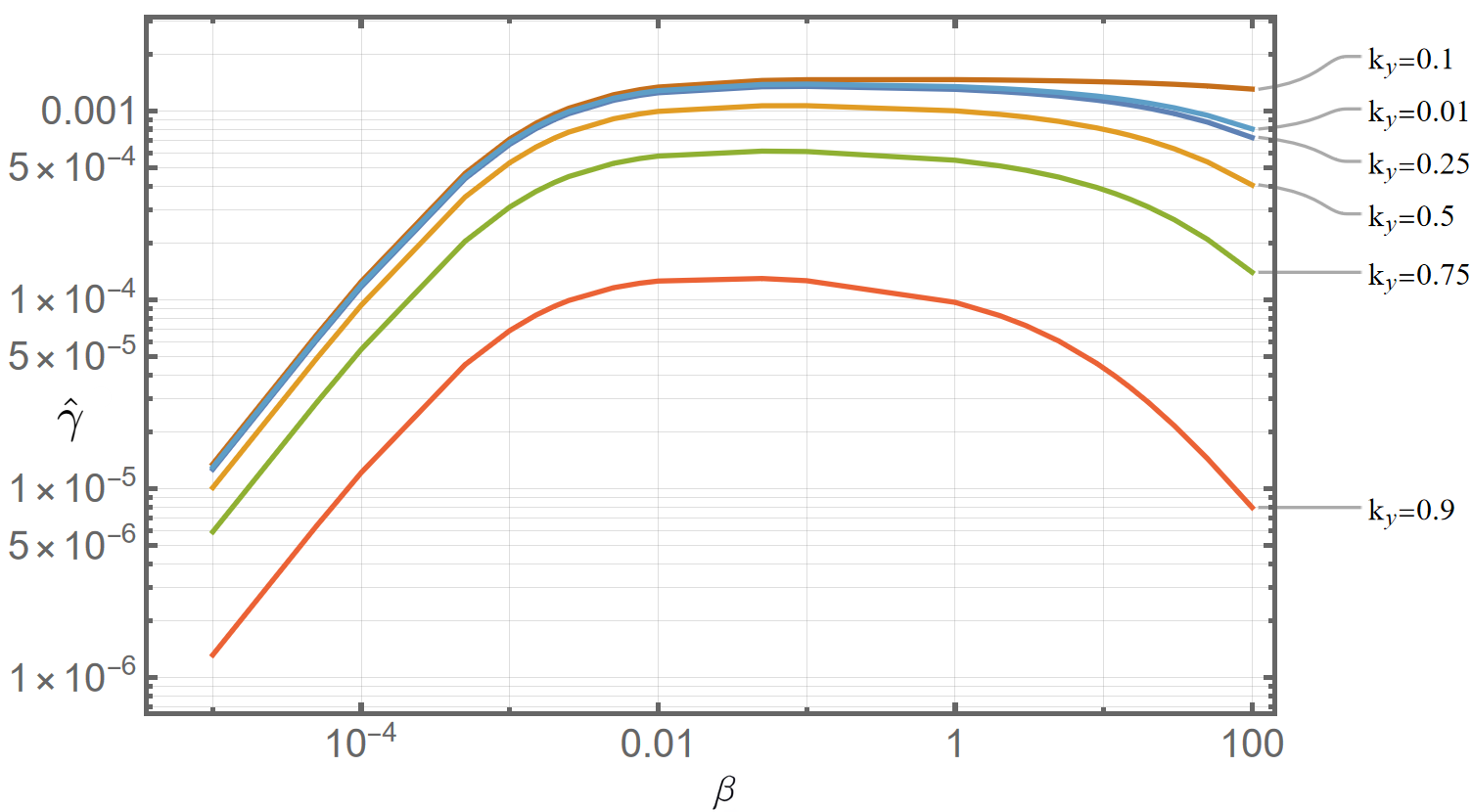}
  \caption{
Analytical estimate for the linear tearing growth rate
$\hat{\gamma}_{\mathrm{lin}}$
at marginal stability as a function of
$\beta$
for different normalized wave numbers
$\hat{k}_{\hat y}$,
obtained from the inversion of
Eq.~(\ref{dispersion_relation}).
For sufficiently small
$\beta \lesssim 10^{-2}$,
the growth rate initially increases with $\beta$,
while for larger values the dependence becomes weaker and
eventually turns over depending on
$\hat{k}_{\hat y}$ and proximity to marginal stability.
}
  \label{gammalin2}
\end{figure}
\end{center}

\begin{center}
\begin{figure}[htp]
\centering
  \includegraphics[trim=0 0 0 0,clip,width=0.45\textwidth]{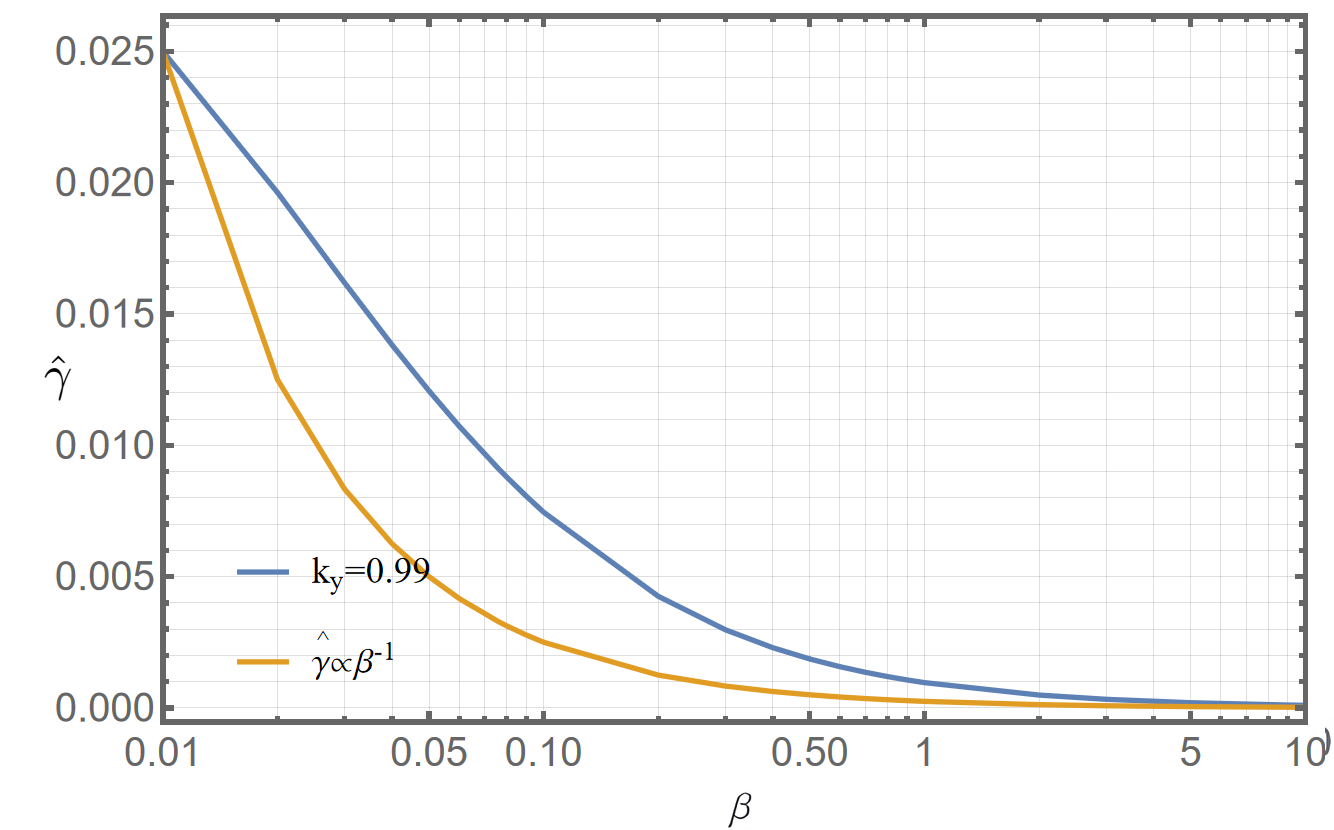}
  \caption{
Analytical estimate for the linear tearing growth rate
$\hat{\gamma}_{\mathrm{lin}}$
at marginal stability over an extended $\beta$ range.
For sufficiently large $\beta$,
the growth rate decreases similar but slower than
$\hat{\gamma}_{\mathrm{lin}} \propto \beta^{-1}$,
consistent with pressure-dominated collisionless reconnection
scalings discussed in
Refs.~\cite{Rogers2007,Pueschel2011}.
}
  \label{gammalin}
\end{figure}
\end{center}

Equation~(\ref{dispersion_relation}) can be inverted
point-wise for given values of
$\hat{k}_{\hat y}$ and $\beta$
in order to obtain the linear tearing growth rate
$\hat{\gamma}_{\mathrm{lin}}$.
The resulting dependence on $\beta$ is shown in
Figs.~\ref{gammalin2}
and~\ref{gammalin}.

For sufficiently small
$\beta \lesssim 10^{-2}$,
the growth rate initially increases with $\beta$.
This behavior originates from the
$\beta^{-1/2}$ contribution in the inner-layer matching
condition of Eq.~(\ref{dispersion_relation}),
which dominates the low-$\beta$ regime.
At larger $\beta$, however, the growth rate decreases
similar but slower than
$\hat{\gamma}_{\mathrm{lin}} \propto \beta^{-1}$,
consistent with pressure-dominated collisionless reconnection
scalings discussed by Rogers et al.~\cite{Rogers2007}
and later gyrokinetic studies
\cite{Pueschel2011}.

The low-$\beta$ increase differs from the trend reported in
three-dimensional gyrofluid simulations by
Biancalani et al.~\cite{Biancalani},
which may be attributed to differences in geometry,
dimensionality, and closure assumptions.
Finally, we emphasize that the present gyrofluid formulation
is derived under a low-$\beta$ ordering. Consequently, the
results presented here should be interpreted within this
regime of validity, and quantitative deviations are expected
as $\beta$ approaches unity.

One has to note that this solution for the growth rate is very much dependent on the equilibrium function $\tilde{A}_{\hat{\parallel},\mathrm{eq}}$, and a deviation has to be expected when e.g.
\begin{align}
a_{\hat{\parallel},eq}(\hat{x})=\frac{1}{ \mathrm{cosh}^2(\hat{x})}
\end{align}
is used. Such initial conditions are very suitable for implementation due to their boundary behavior, and we expect them to behave as analytically predicted in the linear regime due to the similarity in $\hat{\Delta}'$.

To translate the analytical result to the simulations we introduce the box-stability parameter $\hat{\Delta}'_{box}$

\begin{align}
    \hat{\Delta}'_{box} &= \lim_{x\to0^+}\frac{\tilde{A}_{\hat{\parallel},\mathrm{out}}'}{\tilde{A}_{\hat{\parallel},\mathrm{out}}}-\lim_{x\to0^-}\frac{\tilde{A}_{\hat{\parallel},\mathrm{out}}'}{\tilde{A}_{\hat{\parallel},\mathrm{out}}}= 2 \bigg (\frac{1}{\hat{k}}-\hat{k} \bigg ) 
\end{align}
where the Box size enters via $\hat{k}_{\hat{y}}=\pi m/L_{\hat{y}}$. To get e.g. $\hat{k}_{\hat{y}}=1$ with $L_{\hat{y}}=10\pi$ one needs $m=10$, which results in a stability parameter of  $\hat{\Delta}'_{box} = 14.3108$. To obey the limits of the linear stability analysis the system is perturbed with a suitable mode number, close to marginal stability. Note that the aspect ratio $L_{\hat{y}}/L_{\hat{x}}$ does not explicitly show up in the dispersion relation, but as $L_{\hat{y}}$ enters via $\hat{k}_{\hat{y}}$ a change in the aspect ratio results in a different stability parameter. As a result of the small $x$ approximation, rooted in the shrinking automatism of the current sheet, the initial $\hat{k}_{\hat{x}}$ should not influence the general stability of the current sheet, while the aspect ratio of the sheet (note: not the box) still changes. This, and the general stability behavior will be investigated in the following sections with the goal of code verification. A wide scan of aspect-ratio effects is left for later work.

\subsection{Non-modal Analysis}
\label{nomo}

If a system's evolution matrix $\mathcal{A}$ is non-normal, the non-orthogonality of its eigenvectors makes typical spectral analysis less useful \cite{Trefethen,Schmid}. 
Even if all eigenvalues of $\mathcal{A}$ have negative real parts, the evolution operator norm $\lVert \mathrm{e}^{t\mathcal{A}}\rVert$ does not necessarily decay monotonically, but can exhibit transient growth. 
This behavior is related to the energy production operator 
$\mathcal{B} = \mathcal{A}^\dagger + \mathcal{A}$, 
which governs the instantaneous growth rate of the norm.

The evolution matrix
$\mathcal A$
is obtained from the linearized cold-ion system by rewriting
the equations in Fourier space and collecting the coupled
perturbation variables into a state vector
$\hat{\mathbf u}$.
For the constant-gradient equilibrium the resulting operator
has constant coefficients, yielding a linear system of the form
\begin{align}
\partial_t \hat{\mathbf u}
=
\mathcal A\hat{\mathbf u}.
\end{align}
For non-constant equilibrium profiles, the linearized
equations contain products between spatially varying
equilibrium quantities and perturbation fields, for example
terms of the form
$\hat A_{\hat{\parallel},\mathrm{eq}}(\hat x)
\partial_{\hat y}\tilde n_e$
or
$\partial_{\hat x}\hat A_{\hat{\parallel},\mathrm{eq}}
\,\tilde u_e$.
Since multiplication in physical space corresponds to
convolution in Fourier space, these terms couple perturbations
with different $\hat k_{\hat x}$.

As a simple quantification of such behavior, one can analyze the condition number 
\begin{align}
\kappa(V)=\Vert V\Vert \Vert V^{-1}\Vert
\end{align}
with $V$ being a matrix of eigenvectors of the evolution matrix $\mathcal{A}$, for the solution of
\begin{align}
\frac{\partial u(x,t)}{\partial t}= \mathcal{A} u(x,t), 
\quad u(x,t)=\mathrm{e}^{t\mathcal{A}}u(x,0).
\end{align}

For a normal matrix $\mathcal{A}$, the norm of the system evolution satisfies
\begin{align}
\frac{\lVert v(t) \rVert}{\lVert v(0) \rVert} \leq \lVert \mathrm{e}^{\mathcal{A}t} \rVert = \mathrm{e}^{\beta(\mathcal{A}) t},
\end{align}
where $\beta(\mathcal{A}) = \max \{ \mathrm{Re}\,\lambda : \lambda \in \Lambda(\mathcal{A}) \}$ 
denotes the spectral abscissa, with $\Lambda(\mathcal{A})$ being the spectrum of $\mathcal{A}$.
For non-normal matrices, one obtains the bound
\begin{align}
\lVert \mathrm{e}^{\mathcal{A}t} \rVert \le \kappa(V)\,\mathrm{e}^{\beta(\mathcal{A})t},
\end{align}
which can significantly overestimate the actual transient growth.

A sharper characterization of the short-time behavior is given by the numerical range \cite{Trefethen3}

\begin{align}
W(\mathcal A)
=
\left\{
\frac{\langle \mathcal A x,x\rangle}
{\langle x,x\rangle}
:\;
x\neq0
\right\}
\end{align}
and its rightmost point, the numerical abscissa

\begin{align}
\eta(\mathcal A)
=
\sup_{z\in W(\mathcal A)}
\operatorname{Re}(z)
\end{align}
where we make use of the standard Hermitian inner product $\langle \cdot, \cdot \rangle$.
 Another estimate for the transient behavior can be found via the $\epsilon$-pseudospectral abscissa
\begin{align}
\alpha_\epsilon (\mathcal{A}) = \underset{z \in \Lambda_\epsilon (\mathcal{A})}{\mathrm{sup}} Re(z),
\end{align}
which is defined via definition \ref{definits} (see further below). The spectral abscissa $\beta(\mathcal A)$ determines the asymptotic growth or decay rate of the system and is therefore associated with modal stability analysis. 
In contrast, the numerical abscissa $\eta(\mathcal A)$ characterizes the maximal instantaneous growth rate and provides information about possible short-time amplification even when all eigenvalues are stable. 
The $\epsilon$-pseudospectral abscissa $\alpha_\epsilon(\mathcal A)$ measures how far the pseudospectrum extends into the unstable half-plane and therefore quantifies the susceptibility of the system to transient growth and perturbations of the evolution operator. For strongly non-normal operators, $\alpha_\epsilon(\mathcal A)$ can substantially exceed the spectral abscissa, indicating the possibility of significant finite-time amplification despite asymptotic stability. 
When $\mathcal{A}$ is normal, the conditioning number $\kappa$ is 1, while $\kappa=\infty$ in the case of a non diagonalizable matrix. For strongly non-normal operators, quantitative predictions based solely on eigenvalues may become insufficient, since transient amplification can depend sensitively on the conditioning number $\kappa$. Such an analysis is often called "non-modal" and has been carried out in similar context (e.g. Hasegawa-Wakatani model) in various publications \cite{Camargo,Friedman,Held,Trefethen4}. Transient amplification in tearing-type systems has been investigated in several related plasma models
\cite{MacTaggart,Biancalani,Hirota,Ottavini}. A modal analysis of the linearized Full-$F$ model  \cite{LockerThesis} is performed to study any transient behavior. Algebraic manipulations are carried out using Wolfram Mathematica (v.\,12.3) \cite{Mathematica}. First, a simplified case with an initial condition $\hat{A}_{\hat{\parallel,eq}} = a_0 \hat{x}$ and constant gradient $a_0$ is considered, for which $u_e$ and its derivatives vanish. The resulting evolution matrix reads

\begin{equation}
\mathcal{A}_{const.}=
\left(
\begin{array}{cccc}
 a_0 \kappa  & 0 & a_0 \kappa  \hat{k}_{\hat{y}}-\frac{a_0 \kappa }{\alpha } & \frac{a_0 \kappa }{\alpha } \\
 0 & \frac{a_0 \kappa }{\alpha } & \frac{a_0 \kappa  \mu_e}{\alpha }-\frac{a_0 \kappa  \hat{k}_{\hat{y}} \mu_e}{\alpha } & \frac{a_0 \kappa }{\alpha } \\
 -\frac{a_0 \kappa }{\alpha }-\frac{i a_0 \hat{k}_{\hat{y}}}{\hat{k}_{\hat{x}}^2+\hat{k}_{\hat{y}}^2} & \frac{i a_0 \hat{k}_{\hat{y}}}{\hat{k}_{\hat{x}}^2+\hat{k}_{\hat{y}}^2} & 0 & 0 \\
 -\frac{i a_0 \hat{k}_{\hat{y}}}{\mu_e \left(\hat{k}_{\hat{x}}^2+\hat{k}_{\hat{y}}^2\right)} & \frac{i a_0 \hat{k}_{\hat{y}}}{\mu_e \left(\hat{k}_{\hat{x}}^2+\hat{k}_{\hat{y}}^2\right)} & 0 & 0 \\
\end{array}
\right)
\end{equation}
If the initial $\hat{A}_{\hat{\parallel}}$ (and obviously $u_e$ etc.) has a spectrum on its own, one has to carry out the cyclic convolution for the product terms e.g.
\begin{align}
\mathcal{F}(\hat{A}_{{\hat{\parallel}},0}\cdot \frac{\partial}{\partial y}\hat{n}_e)=(2 \pi)^{-\frac{n}{2}} \mathcal{F}(\hat{A}_{{\hat{\parallel}},0})* \mathcal{F}(\frac{\partial}{\partial y}\hat{n}_e).
\end{align}
Again, the evolution depends on the initial vector potential $\hat{A}_{{\hat{\parallel}},0}$ (determining all the dependent quantities) and one option is to make use of the Fourier-sum representation, singling out one mode number, and further implying periodic boundary conditions, so
\begin{align}
\mathcal{F}(\hat{A}_{{\hat{\parallel}},0})=\mathcal{F}(a_0\cos (nx))=a_0 \sqrt{\frac{\pi}{2}}(\delta (k-n)+\delta (n+k)).
\end{align}
One can finally use the sifting property of the convolution 
\begin{align} 
\int_\infty^{-\infty} f(x)\delta(x-n) dx=f(n)
\end{align}
to obtain an evolution matrix for each mode number $n$. The equilibrium profile introduces discrete spectral support at 
$\hat{k}_{\hat{x}}=\pm n$. 
Consequently, the convolution terms act as mode-coupling operators in Fourier space, shifting perturbation modes according to 
$\hat{k}_{\hat{x}}\rightarrow \hat{k}_{\hat{x}}\pm n$. 
The resulting evolution matrix therefore remains parametrized by the perturbation wave numbers $(\hat{k}_{\hat{x}},\hat{k}_{\hat{y}})$, while the integer $n$ labels the (constant) equilibrium spectral component.

\begin{widetext}
\begin{equation}
\mathcal{A}_n=
\left(
\begin{array}{cccc}
 \frac{\beta  n \hat{k}_{\hat{y}} \left(\mu _e \left(2 \hat{k}_{\hat{x}}^2+2 \hat{k}_{\hat{y}}^2-\sqrt{2 \pi } n^4-2\right)+\sqrt{2 \pi } n^2+2\right)}{4 \mu _e \left(\hat{k}_{\hat{x}}^2+\hat{k}_{\hat{y}}^2-1\right)+4} & 0 & \frac{\beta  n \mu _e \hat{k}_{\hat{y}} \left(\hat{k}_{\hat{x}}^2+\hat{k}_{\hat{y}}^2+n^2-1\right)}{2 \mu _e
   \left(\hat{k}_{\hat{x}}^2+\hat{k}_{\hat{y}}^2-1\right)+2} & \frac{\beta  n \hat{k}_{\hat{y}} \left(1-n^2 \mu _e\right)}{2 \mu _e \left(\hat{k}_{\hat{x}}^2+\hat{k}_{\hat{y}}^2-1\right)+2} \\
 \frac{\sqrt{\frac{\pi }{2}} \beta  n^3 \mu _e \hat{k}_{\hat{y}}}{2 \mu _e \left(\hat{k}_{\hat{x}}^2+\hat{k}_{\hat{y}}^2-1\right)+2} & \frac{1}{2} \beta  n \hat{k}_{\hat{y}} & -\frac{\beta  n \mu _e \hat{k}_{\hat{y}}}{2 \mu _e \left(\hat{k}_{\hat{x}}^2+\hat{k}_{\hat{y}}^2-1\right)+2} & \frac{1}{2} \beta  n \hat{k}_{\hat{y}} \left(\frac{\mu _e}{\mu _e
   \left(\hat{k}_{\hat{x}}^2+\hat{k}_{\hat{y}}^2-1\right)+1}+1\right) \\
 \frac{1}{4} \left((2 \pi  \beta -i) \mu _e+\beta  \left(\frac{4 \pi }{\mu _e+1}-4 \pi +2\right)+i\right) & \frac{1}{4} i \left(\mu _e-1\right) & -\frac{\sqrt{\frac{\pi }{2}} \beta  \mu _e}{\mu _e+1} & \frac{\sqrt{\frac{\pi }{2}} \beta 
   \left(\mu _e-1\right) \mu _e}{2 \left(\mu _e+1\right)} \\
 -\frac{i \hat{k}_{\hat{y}}}{2 \left(\hat{k}_{\hat{x}}^2+\hat{k}_{\hat{y}}^2\right)} & -\frac{i \hat{k}_{\hat{y}}}{2 \left(\hat{k}_{\hat{x}}^2+\hat{k}_{\hat{y}}^2\right)} & 0 & 0 \\
\end{array}
\right)
\label{eq:Amsystem}
\end{equation}
\end{widetext}

The resulting eigensystem shown in eq. \ref{eq:Amsystem} can then be analyzed for different initialisations with mode number $m$, varying $\beta$ and $\hat{k}_{\hat{x},\hat{y}}$. At this point it makes sense to single out a low $\hat{k}_{\hat{y}}$ and compare a few different values to investigate how the condition number scales with the system parameters. For $\hat{k}_{\hat{y}} = 0$, the condition number remains approximately constant at $\kappa=10^6$ for all parameters, and generally large for $\hat{k}_{\hat{y}} \neq 0$ with a local minimum around marginal stability, yielding a possible explanation for transient behavior.

To analyze the spectrum of the evolution operator $\mathcal{A}$, we make use of the resolvent of $\mathcal{A}\in \mathbb{C}^{n\times n}$
\begin{align}
\mathcal{R}(\mathcal{A})=(z\mathbb{I}_{n}-\mathcal{A})^{-1}
\end{align} 
which is obviously singular when $z$ is an eigenvalue $\lambda_\mathcal{A}$ of the operator. In the same way, we can define the spectrum of the operator $\Lambda (\mathcal{A})$ as the set of all values for which the resolvent becomes singular. Analogously one can define the $\epsilon$ pseudospectrum $\Lambda_{\epsilon} (\mathcal{A})$ as the set of complex numbers that satisfy definition \ref{definits} \cite{Trefethen}.

\begin{center}
\begin{figure}[htp]
	\centering
\includegraphics[trim=0 0 0 0,clip,width=0.49\textwidth]{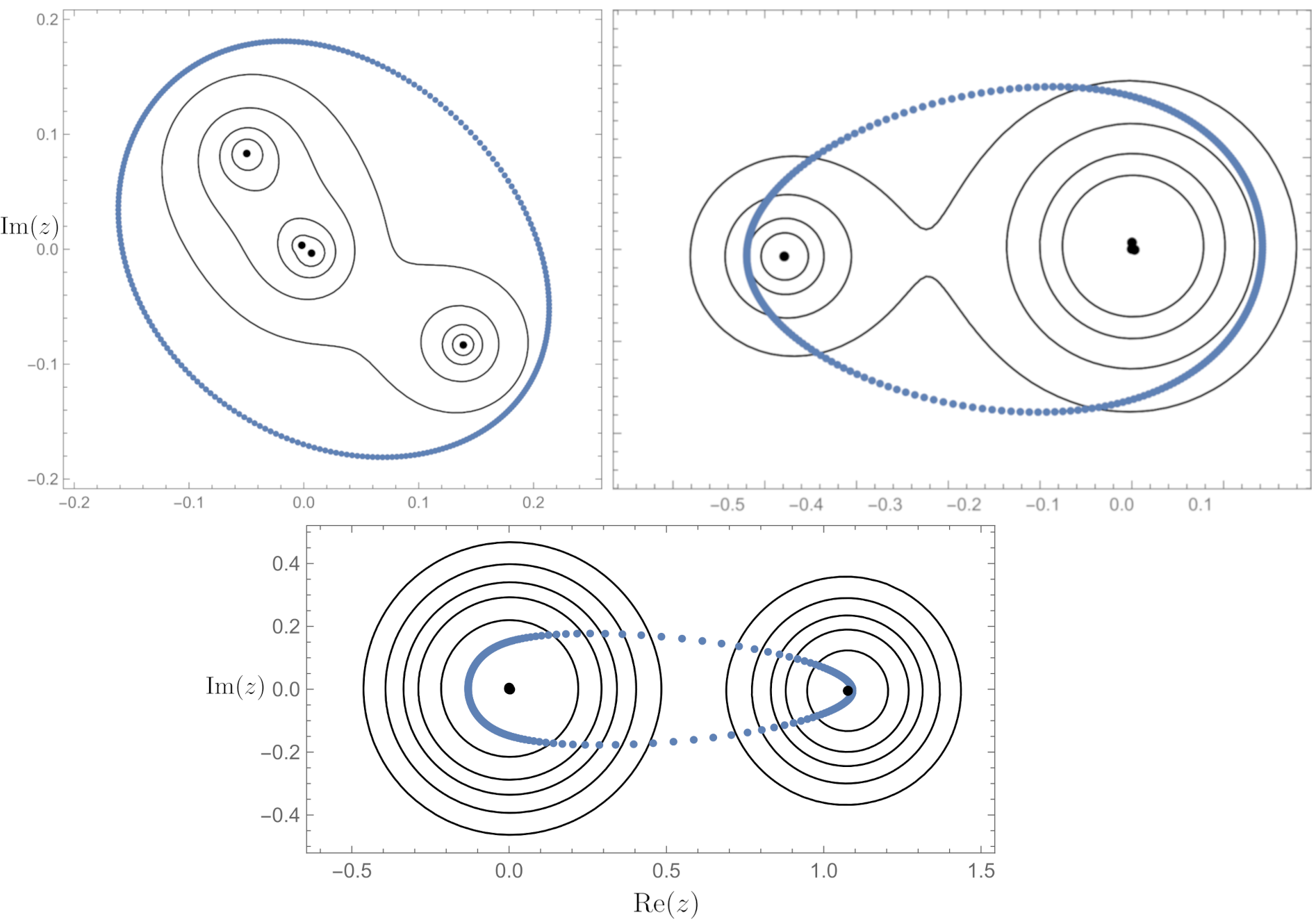}
\caption{Black contours show $\epsilon$-pseudospectral level sets of the evolution operator for decreasing perturbation levels $\epsilon = 10^{-1},\,10^{-2},\,10^{-3},\,10^{-4}$, while black dots denote eigenvalues. 
The blue dotted curve indicates the boundary of the numerical range $W(\mathcal A)$, whose rightmost point determines the numerical abscissa $\eta(\mathcal A)$. 
Contour plots of the $\epsilon$-pseudospectrum of $\mathcal{A}_n$ are shown for $n=1$ and $\hat{k}_{\hat{y}} = 0.1,\,1,\,10$ (left, right, bottom), together with the numerical range boundary. 
For small $\hat{k}_{\hat{y}}$, the pseudospectral contours exhibit strong deformation and merging, indicating pronounced non-normality and the possibility of significant transient amplification. 
As $\hat{k}_{\hat{y}}$ increases, the contours become progressively more localized and approach the circular structure characteristic of nearly normal operators. 
Simultaneously, the dominant eigenvalues approach the real axis and shift progressively toward the unstable half-plane.}
  \label{numericalabscissa}
\end{figure}
\end{center}

\begin{definition}[$\epsilon$- pseudospectrum]\label{definits} The set of complex numbers $z$, that satisfy
\begin{itemize} 
\item[(i)]$\lVert (z\mathbb{I}_{n}-\mathcal{A})^{-1}  \rVert \geq \epsilon^{-1}$,
\item[(ii)]$\sigma_{min}(z\mathbb{I}_{n}-\mathcal{A})\leq \epsilon $, where $\sigma_{min}$ denotes the smallest singular value,
\item[(iii)]$\lVert \mathcal{A}u-zu  \rVert \leq \epsilon$ for some vector $u$ with $\lVert u\rVert =1$,
\item[(iv)]$z$ is an eigenvalue of $\mathcal{A}+E$, where $E$ is a perturbation to $\mathcal{A}$ with $\lVert E \rVert \leq \epsilon$.
\end{itemize}
is called, the $epsilon$-pseudospectrum of Operator $A$.
\end{definition}
In a practical sense, one can now plot $\epsilon$-contour lines of $(z\mathbb{I}_{n}-\mathcal{A})$, which, in terms of the spectrum, means, that we can look at the subsets of "almost" singular values, and this $\epsilon$-pseudospectrum can be viewed as the spectrum of some perturbed matrix. Nevertheless, the pseudospectrum does not only contain information about a change in the spectrum under a perturbation of the operator but also gives information about the operator itself \cite{Trefethen2}. The pseudospectrum can be used to gain insight on transient effects on instabilities, and further, one can try to find parameters for the operator to prevent such transient behavior. In Fig. \ref{numericalabscissa}, the $\epsilon$-pseudospectrum of $\mathcal{A}$ is shown together with the boundary of the numerical range (blue dotted line) for an initial condition with $n=1$. 
The rightmost point of the numerical range determines the numerical abscissa and provides an upper bound for the instantaneous growth rate at $t=0$. 
For small $\hat{k}_{\hat{y}}$, the pseudospectral contours exhibit strong deformation and merging, indicating pronounced non-normality and the possibility of significant transient amplification. 
As $\hat{k}_{\hat{y}}$ increases, the contours become progressively more localized and approach the circular structure characteristic of nearly normal operators. 
Simultaneously, the dominant eigenvalues approach the real axis and shift toward the unstable half-plane. 
The $\epsilon$-pseudospectra extend substantially beyond the eigenvalue spectrum, indicating strong sensitivity of the evolution operator to perturbations and the presence of transient growth. 
The numerical abscissa remains positive for all considered $\hat{k}_{\hat{y}}$, implying the existence of initial conditions with short-time exponential amplification. 
From the extent of the $\epsilon$-pseudospectra, transient growth factors of order $1/\epsilon$ are expected. 
This indicates the possibility of substantial transient amplification and provides a possible mechanism facilitating the transition toward rapid nonlinear reconnection dynamics.

One may estimate transient growth by sampling perturbation matrices $E$ with $\lVert E \rVert = \epsilon \to 0$ and evaluating the corresponding spectral abscissa \cite{Trefethen},
\begin{align}
\gamma(A) = \lim_{\epsilon \to 0} \alpha_\epsilon(A).
\end{align}
Equivalently, one may infer bounds on transient growth directly from the $\epsilon$-pseudospectrum by considering the rightmost extent of $\Lambda_\epsilon(A)$ in the complex plane. For $\hat{k}_{\hat{y}} \leq 1$, the largest observed values of $\alpha_\epsilon(A)$ are approximately $0.2$, which provides a more realistic upper bound on the transient growth rate than that suggested by the condition number alone.  

In the present analytical estimate, a very simple class of initial conditions has been employed. In practice, however, the dynamics may involve more complex perturbations, and it is therefore not surprising if estimates based on a restricted set of modes (e.g.\ a fixed mode number) do not quantitatively match numerical simulations.  

We conclude that the linearized operator exhibits very large condition numbers and is therefore strongly non-normal. 
This implies that transient amplification is a natural feature of the linearized dynamics and may occur in addition to, or before, the asymptotic modal growth described by the eigenvalues. 
The non-modal analysis should therefore not be interpreted as a direct description of the nonlinear explosive phase itself. 
Rather, it shows that the underlying Full-$F$ operator is susceptible to transient amplification, which can enhance perturbation amplitudes and thereby facilitate the transition toward nonlinear explosive reconnection\cite{Aydemir,Biancalani,Comisso,Stanier,Granier,Ottavini}. 
Since the nonlinear evolution continuously modifies the effective operator around the instantaneous state, non-modal amplification and nonlinear explosive growth should not be regarded as completely separate mechanisms.

The dominant contribution to non-normality arises from the
off-diagonal coupling between the electromagnetic variables
and the density and velocity perturbations, in particular from
terms proportional to the equilibrium magnetic shear
$\hat{A}_{\parallel,\mathrm{eq}}'$
and to electron inertia through $\mu_e$.
These couplings render the evolution matrix
$\mathcal A$
non-self-adjoint and couple perturbations associated with
different physical energy channels.
For the Fourier-sum equilibrium,
additional non-normality is introduced by convolution-induced
mode coupling, which shifts perturbations according to
$\hat{k}_{\hat{x}} \rightarrow \hat{k}_{\hat{x}} \pm n$.
Consequently, transient amplification may occur even in
parameter regimes where the asymptotic modal growth rate is
small.

\subsection{Tearing Instability and Rapid Nonlinear Acceleration of Magnetic Reconnection}\label{tearing}

We now turn from the linearized stability analysis to
fully nonlinear simulations of a tearing-unstable Harris-type
sheet \cite{porcelli2002recent}.
\begin{gather}\label{harrinit}
\hat{A}_{\hat{\parallel}}=\beta\cdot \bigg(\frac{A_0}{\mathrm{cosh}(\eta \cdot 4 \pi  \hat{x}/L_{\hat{x}})^2}+ A_1\cdot \mathrm{cos}(m \cdot 2 \pi \hat{y}/L_{\hat{y}})  \bigg )\mathrm{cos}(\pi \hat{x}/L_{\hat{x}})\\
\hat{u}_e=\frac{1}{N_0\cdot\beta}\hat{\Delta} \hat{A}_{\hat{\parallel}} 
\end{gather}

with the parameters: sheet amplitude := $A_0$, perturbation amplitude := $A_1$, perturbation mode number (in $y-$direction) := $m$, background density := $N_0$, numerical sheet width parameter := $\eta$. Throughout the following, the term
"explosive" refers to the rapid nonlinear acceleration of
the reconnection dynamics following the initial
approximately linear tearing phase. To capture all X-points, one can make use of the relative global partition reconnection rate \cite{Priest}
\begin{align}\label{defrateeq}
	\hat{\gamma} =\frac{d}{d\hat{t}} \mathrm{ln}(\sum_{x_p}\Delta \hat{\Psi}(\hat t))=\frac{d}{d\hat{t}}\mathrm{ln}(\sum_{x_p} |\hat A_{\hat{\parallel},\mathrm{eq}}(\mathbf{x}_{0},\hat t)
-
\hat A_{\hat{\parallel}}(\mathbf{x}_{p},\hat t)| )
\end{align}
where all the X-points are located by finding local minima and tracking their evolution and $\hat{\Delta} \hat{\Psi}=|\hat{A}_{\hat{\parallel}}(\vec{x},t)-\hat{A}_{\hat{\parallel}}(\vec{x}_0,0)|$. This method requires the choice of an equilibrium vector potential
$A_{\hat{\parallel},\mathrm{eq}}$, which for the analytical Harris-sheet
is given by Eq.~(\ref{harrinit}). Since the Harris-sheet equilibrium
depends only on the radial coordinate $\hat{x}$, it is entirely
contained in the $\hat{k}_{\hat{y}}=0$ Fourier component. In the
numerical simulation, however, the initialized equilibrium is
represented only through its discrete approximation, which is
subject to spectral discretization, numerical filtering, boundary
treatment, and finite floating-point precision. Comparing the
perturbation directly with the analytical equilibrium would therefore
attribute these small numerical differences to the perturbation.
Instead, we define the numerical equilibrium as the
$\hat{k}_{\hat{y}}=0$ Fourier component of
$\hat{A}_{\hat{\parallel}}$, providing a consistent separation
between the background field and the finite-$\hat{k}_{\hat{y}}$
perturbations throughout the simulation. The resulting differences
with respect to the analytical equilibrium are small and depend on
the resolved $\hat{k}_{\hat{x}}$ spectrum \cite{ScottV2}.

For the scope of this paper, the numerical sheet-width parameter
$\eta$ was set to unity, resulting in a physical sheet width of
\begin{align}
\frac{L_{\hat x}}{4\pi\eta}
=
\frac{32\rho_s}{4\pi}
=
\frac{8}{\pi}\rho_s
\approx2.55\rho_s
\end{align}
for a domain of $L_{\hat{x}}=32\rho_s$.

\begin{figure}[htp]
\centering
\includegraphics[trim=0 0 0 0,clip,width=0.49\textwidth]
{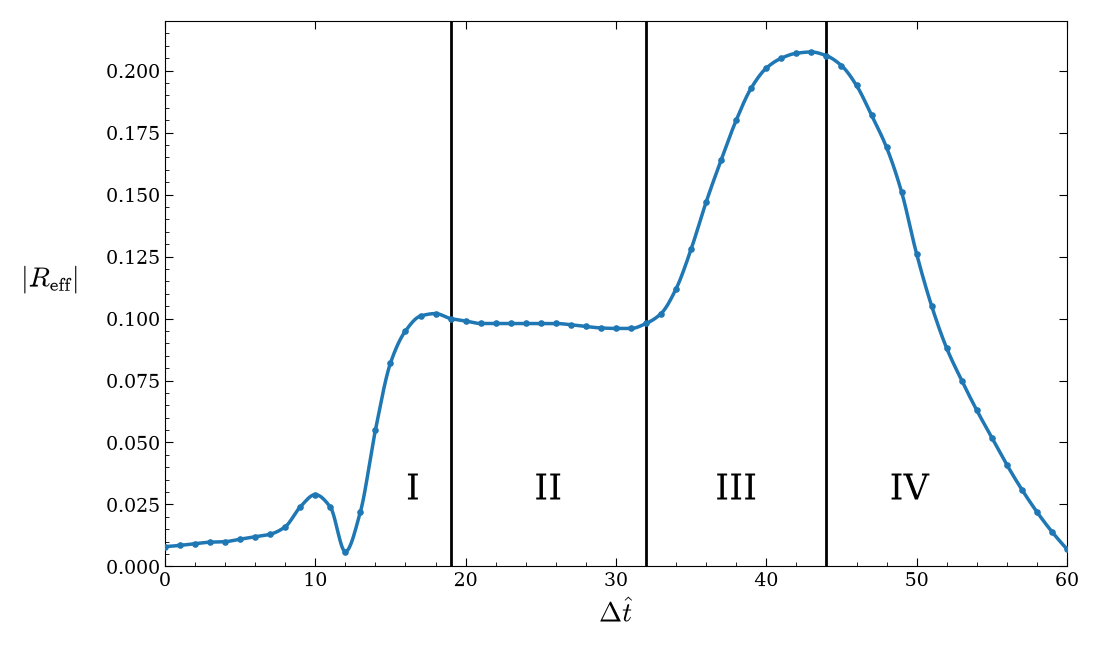}
\caption{
Normalized reconnection rate
$R_{\mathrm{rec}}=E_{\mathrm{rec}}/(B_0v_A)$
for a warm-ion Harris sheet with
$\tau_i=1$, $\beta=5\times10^{-3}$, and $m=1$.
The horizontal axis is given in output intervals
$\Delta\hat t=\hat t\,\Delta_{\mathrm{output}}$
with
$\Delta_{\mathrm{output}}=1000$.
The evolution proceeds through an initial transient stage (I),
a quasi-steady reconnection stage (II), a phase of rapid
nonlinear acceleration (III), and final saturation (IV).
This simulation was selected because these phases are
particularly well separated.
During stage II,
$R_{\mathrm{rec}}\approx 0.1$,
while stage III reaches a peak value of approximately
$R_{\mathrm{rec}}\approx 0.2$.
}
\label{explrecon}
\end{figure}

Figure~\ref{explrecon} shows the nonlinear evolution of a
tearing-unstable Harris-sheet configuration. The evolution
proceeds through several distinct stages, including an initial
transient phase, a quasi-steady reconnection phase, a subsequent
phase of rapid nonlinear acceleration, and final saturation
associated with the topological rearrangement of the magnetic
field.

Since the normalized vector potential evolved by the model
contains the absorbed beta factor,
$\hat A_{\hat{\parallel}}=\beta\widetilde A_{\hat{\parallel}}$,
the normalized reconnected magnetic flux is defined as
\begin{align}
\hat{\Psi}(\hat t)
=
\hat A_{\hat{\parallel}}(\mathbf{x}_{0},\hat t)
-
\hat A_{\hat{\parallel}}(\mathbf{x}_{X},\hat t),
\end{align}
where $\mathbf{x}_{0}$ and $\mathbf{x}_{X}$ denote the
corresponding O- and X-points.
Using
$A_{\parallel}=B_0\rho_s\hat A_{\parallel}/\beta$
and
$t=(L_\perp/c_s)\hat t$,
the normalized reconnection rate becomes
\begin{align}
R_{\mathrm{rec}}
=
\frac{E_{\mathrm{rec}}}{B_0v_A}
=
\frac{\delta}{\sqrt{\beta}}
\left|
\frac{d\hat{\Psi}}{d\hat t}
\right|,
\end{align}
where
$\delta=\rho_s/L_\perp$.
Since $\delta=1$ throughout the present work,
\begin{align}
R_{\mathrm{rec}}
=
\frac{1}{\sqrt{\beta}}
\left|
\frac{d\hat{\Psi}}{d\hat t}
\right|.
\end{align}

This diagnostic is distinct from the logarithmic tearing
growth rate for a single X-point (eq. \ref{defrateeq})
\begin{align}
\hat{\gamma}
=
\frac{1}{\Delta\hat{\Psi}}
\frac{d\,\Delta\hat{\Psi}}{d\hat t}
=
\frac{d}{d\hat t}
\ln\!\left(\Delta\hat{\Psi}\right).
\end{align}

which characterizes the relative growth of the reconnected
flux rather than the reconnection electric field. Since both diagnostics are derived from the same reconnected
flux, they may exhibit similar temporal features, although
they differ by the instantaneous factor
$1/\Delta\hat{\Psi}$.

For the case shown in Fig.~\ref{explrecon},
the normalized reconnection rate remains of order
$R_{\mathrm{rec}}\approx 0.1$
during the quasi-steady phase and reaches a peak value of
approximately
$R_{\mathrm{rec}}\approx 0.2$
during the subsequent nonlinear acceleration phase,
consistent with previously reported fast collisionless
reconnection rates
\cite{Shay1999,Birn2001}.
The simulation shown in Fig.~\ref{explrecon} was selected
because it exhibits particularly well-separated dynamical
phases, allowing the transition from quasi-steady to rapid
nonlinear reconnection to be identified clearly. Depending on
the parameters, however, the evolution may become more complex,
with overlapping linear and nonlinear phases and transient
behavior arising from the interaction of multiple unstable
modes.

\begin{figure}[htp]
\centering
  \includegraphics[trim=0 0 0 0,clip,width=0.49\textwidth]{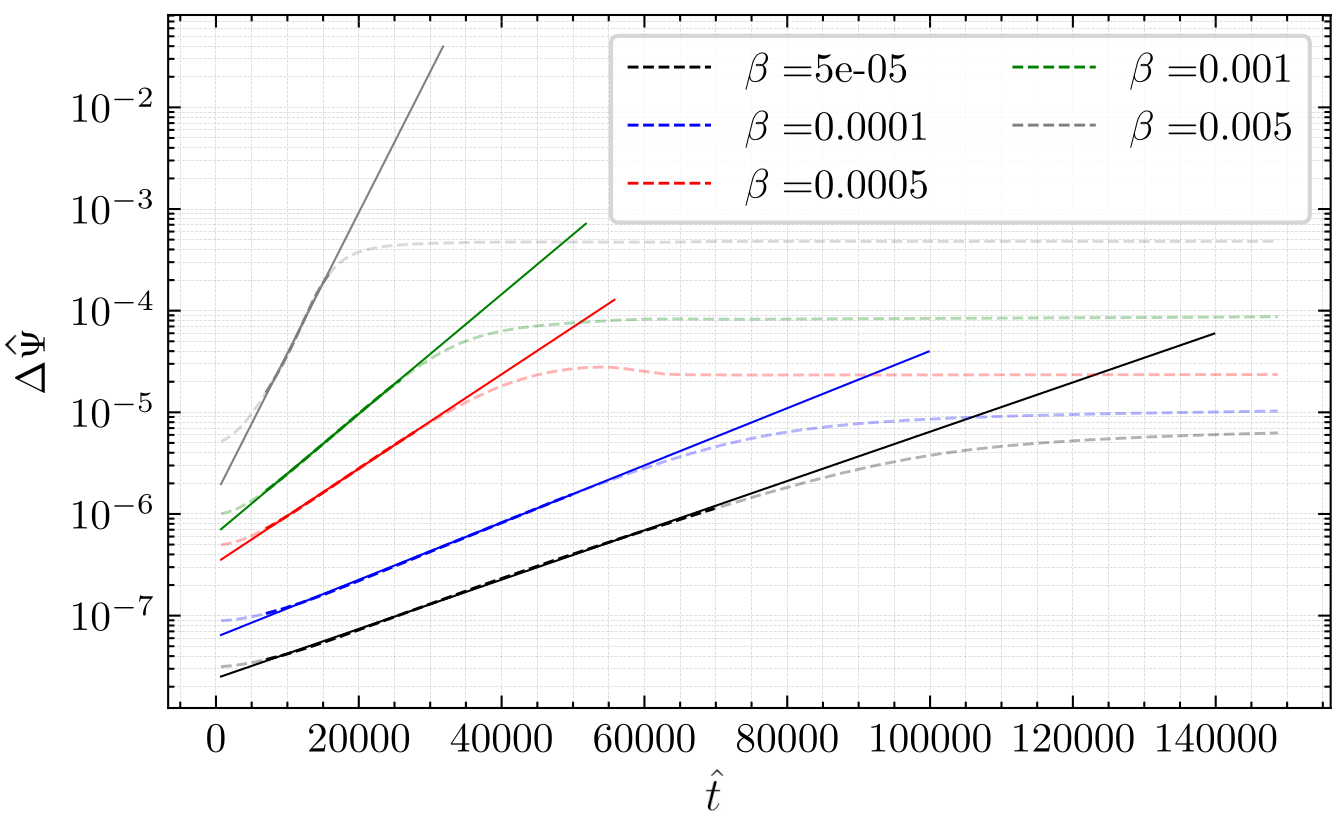}
\caption{
Logarithmic time evolution of $\hat{\Delta}\hat{\Psi}$ for different values of $\beta$. As expected from the dispersion relation, the growth rate decreases with decreasing $\beta$ (corresponding to increasing $\hat d_e$). The approximately linear sections indicate the exponential-growth phase.
}
  \label{Delapslin}
\end{figure}

In the $m=1$ case, reconnection proceeds through a single
X-point, making it easy to distinguish the different phases
of the evolution. For decreasing $\beta$ (corresponding to increasing $\hat d_e$), the obtained
growth rates decrease and generally exhibit a phase of
exponential growth. The simulations shown in
Fig.~\ref{Delapslin} correspond to a separate parameter scan
and are not identical to the illustrative case presented in
Fig.~\ref{explrecon}. The presence of the linear phase is
particularly visible in the logarithmic evolution of
$\hat{\Delta}\hat{\Psi}$ shown in
Fig.~\ref{Delapslin}.

Finally, the growth rate is investigated in the regime where
the analytical estimate is applicable, using the small-stability-
parameter approximation
$\hat{\Delta}'\ll1$
with
$m=4.9$
for
$\tau_i\in\{0,1\}$.
This regime is close to marginal stability and is obtained by
choosing
$\hat{k}_{\hat y}$
close to, but smaller than, unity.
The linear phases of a current sheet perturbed with
$m=1$,
for which the analytical estimate is not expected to be
quantitatively valid, are additionally compared for arbitrary-
wavelength and long-wavelength-approximated (LWL)
polarization at
$L_{\hat{x}}=L_{\hat{y}}=32\rho_s$,
$n_{\hat{x}}=n_{\hat{y}}=1024$,
and
$\nu=10^{-8}$.

\begin{figure*}[htp]
\centering
\includegraphics[trim=0 0 0 0,clip,width=0.9\textwidth]
{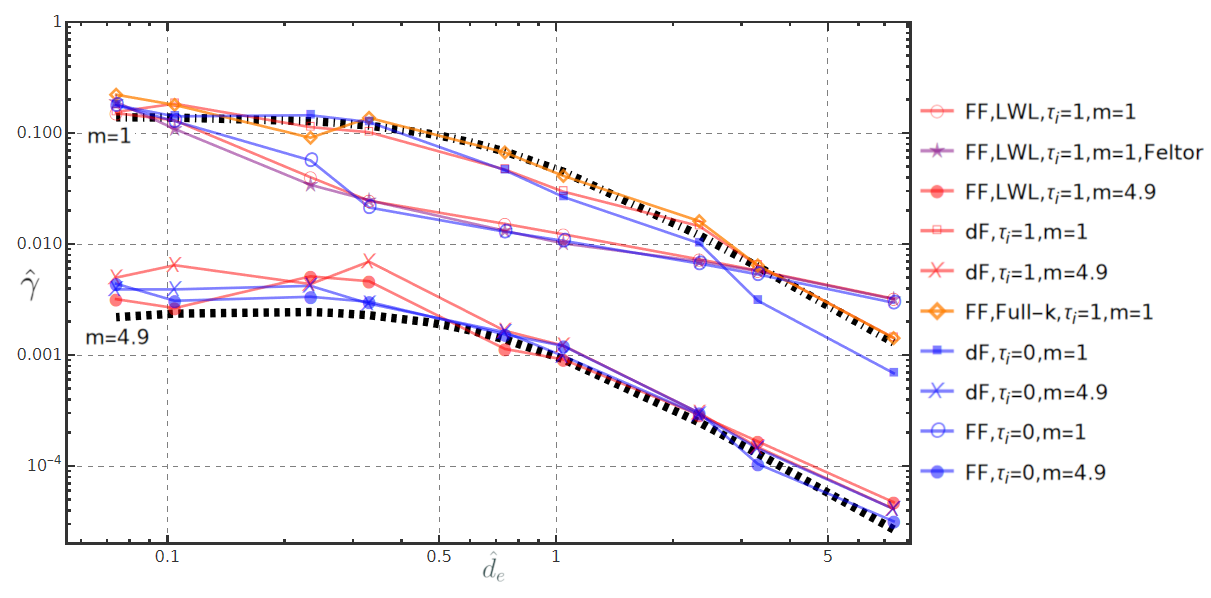}
\caption{
Analytical estimates
$\hat{\gamma}_{\mathrm{lin}}$
(black dotted and dot-dashed lines) and numerically obtained
linear growth rates
$\hat{\gamma}$
for marginally stable
$m=4.9$
and unstable
$m=1$
perturbations using different
$\delta F$
and Full-$F$ models.
The marginally stable cases show good agreement with the
analytical estimate, whereas the
$m=1$
results are included only for qualitative comparison.
For
$m=1$,
deviations become visible already at intermediate
$\hat d_e$
and increase as
$\hat d_e$
approaches unity.
The closest agreement is recovered when arbitrary-wavelength
polarization is combined with weak hyperviscosity
$\nu\sim10^{-8}$.
Ion-FLR effects remain comparatively small in the square
domain for most of the considered cases.
}
\label{Gammasneu}
\end{figure*}

In Fig.~\ref{Gammasneu}, the numerically obtained linear
growth rates are compared with the analytical estimate for
different parameters.
Good agreement is observed for the marginally stable cases,
whereas for
$m=1$
deviations become visible already at intermediate values of
$\hat d_e$
and grow more pronounced as
$\hat d_e$
approaches unity.
The
$m=1$
case is also compared with results obtained from the
open-source gyrofluid code FELTOR~\cite{Feltor}.
As expected, FLR effects play only a minor role in most cases,
particularly when the polarization equation is used in
long-wavelength-approximated form.
Sensitivity to the polarization model is nevertheless observed
for the
$m=1$
system, while the deviation from the analytical estimate is
reduced when arbitrary-wavelength polarization is combined
with weak hyperviscosity
$\nu\sim10^{-8}$.

We emphasize that the analytical estimate was derived in the
cold-ion limit
$\tau_i\rightarrow0$,
whereas some of the presented Full-$F$, Full-$k$ simulations
employ finite ion temperature with
$\tau_i=1$.
The comparison should therefore be interpreted as qualitative
rather than as an exact quantitative agreement.
The stronger departure of the LWL model becomes increasingly
apparent as
$\hat d_e=d_e/\rho_s$
increases and finite-wavelength effects become relevant.
Similar sensitivity of linear gyrofluid dynamics to the
Oberbeck--Boussinesq approximation and polarization modeling
has been discussed by Held et al.~\cite{Held}.

\begin{figure*}[htp]
\centering
  \includegraphics[trim=0 0 0 0,clip,width=0.95\textwidth]{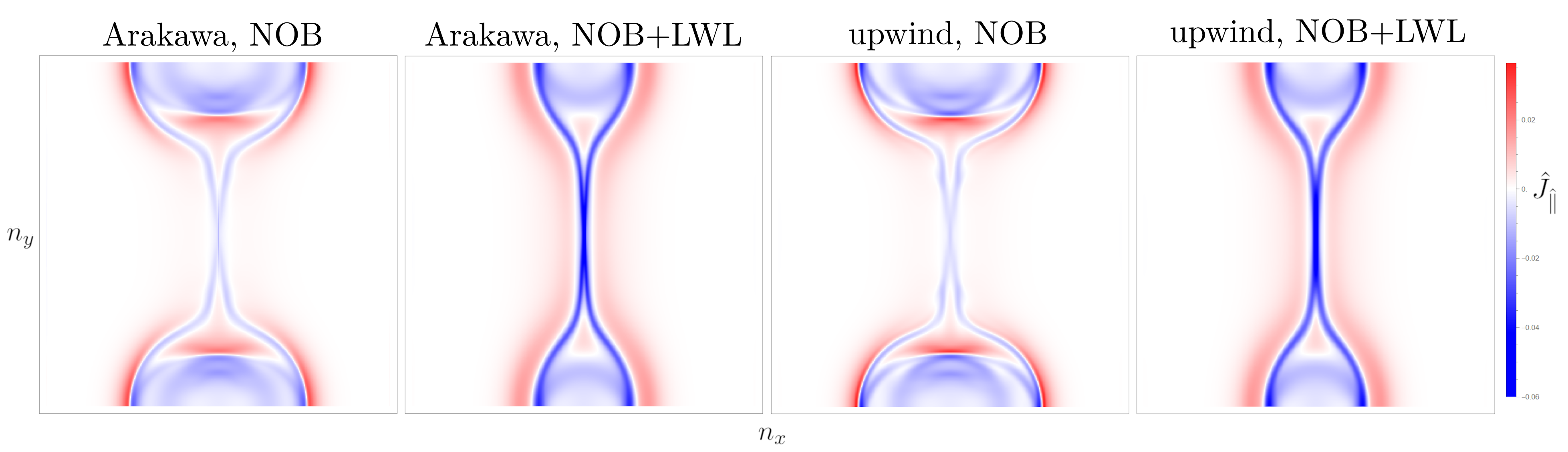}
 \caption{
 Comparison of the Arakawa discretization with
 $\nu=10^{-7}$ and an upwind scheme for the
 arbitrary-wavelength NOB and NOB+LWL polarization models at
 comparable times. Both discretization schemes yield similar
 current-sheet structures, while reconnection proceeds more
 rapidly for the arbitrary-wavelength NOB model than for the
 NOB+LWL approximation.
 }
  \label{Upwind}
\end{figure*}

Figure~\ref{Upwind} compares the Arakawa discretization with
$\nu = 10^{-7}$ and an upwind scheme for the arbitrary-wavelength
NOB model and its long-wavelength-approximated counterpart
(NOB+LWL). The latter discretization introduces numerical
dissipation on its own~\cite{hirsch2007numerical}. The color
contours represent the parallel current density
$\hat{J}_{\hat{\parallel}}$, illustrating the formation and
nonlinear evolution of the reconnecting current sheet. Both
discretization schemes produce qualitatively and quantitatively
similar current structures, indicating that the centered
Arakawa scheme with weak hyperdiffusion provides sufficient
regularization without introducing significant numerical
artifacts. Moreover, reconnection proceeds more rapidly for the
arbitrary-wavelength NOB polarization than for the NOB+LWL
approximation, demonstrating the influence of finite-wavelength
polarization effects. Overall, the agreement between the
Arakawa and upwind results confirms that the observed
differences arise from the physical model rather than from the
numerical discretization.

\begin{figure*}[htp]
\centering
  \includegraphics[trim=0 0 0 0,clip,width=1.\textwidth]{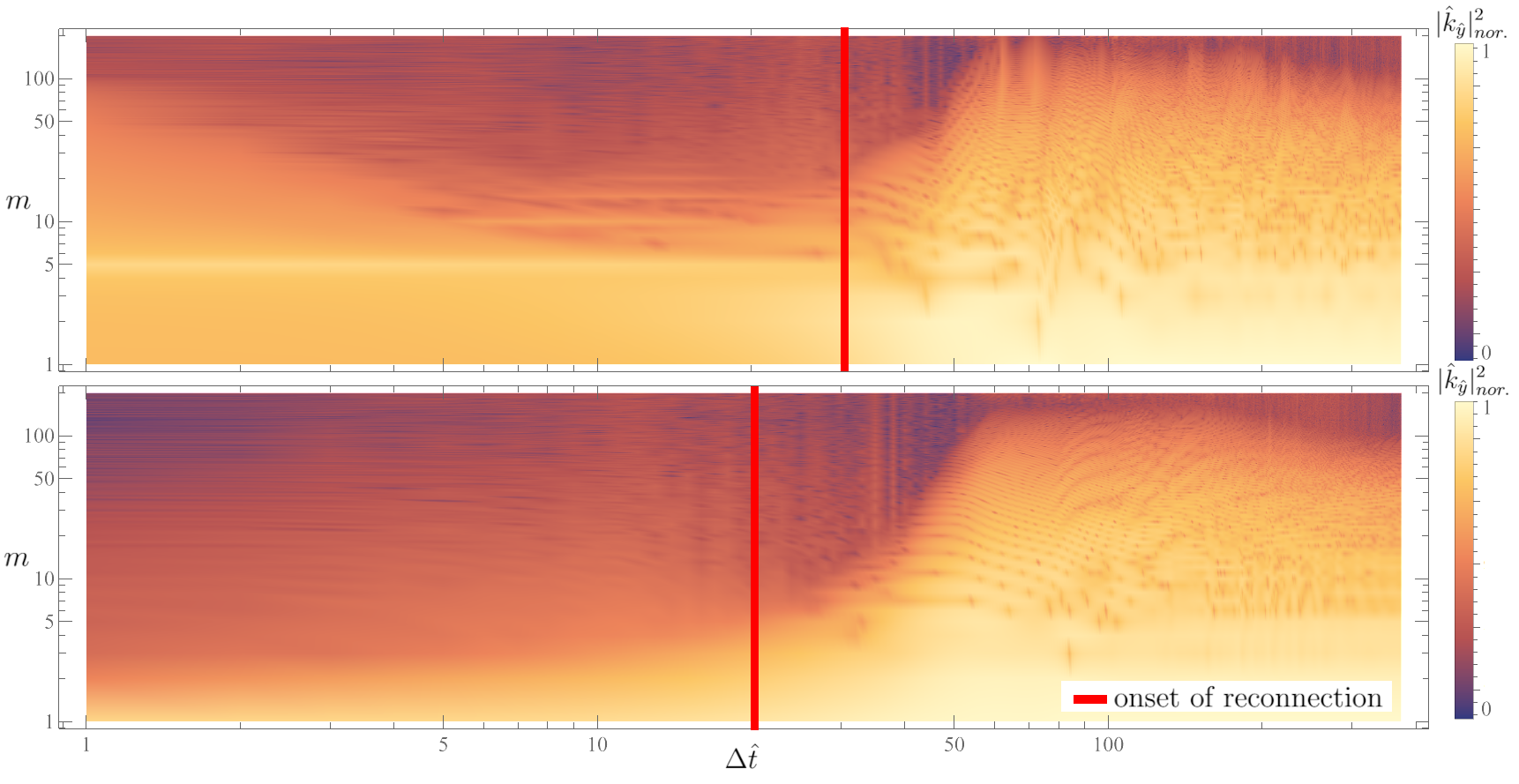}
\caption{
Time evolution of the normalized
$\hat{k}_{\hat y}$ spectrum of the parallel magnetic vector
potential $\hat{A}_{\hat{\parallel}}$ for the marginally stable
$m=4.9$ case (left) and the unstable $m=1$ case (right). The
equilibrium mode $\hat{k}_{\hat y}=0$ is omitted from the
displayed spectrum. The $m=1$ panel corresponds to the same
simulation shown in Fig.~\ref{explrecon}. The spreading of
spectral energy away from the initially excited mode
illustrates nonlinear mode coupling during reconnection.
}
  \label{spectra}
\end{figure*}

\noindent Figure~\ref{spectra} shows the time evolution of the normalized
$\hat{k}_{\hat y}$ spectrum of the parallel magnetic vector
potential $\hat{A}_{\hat{\parallel}}$ for a marginally stable
$m=4.9$ current sheet and an unstable $m=1$ current sheet.

For the marginally stable case, the initially excited mode
persists for an extended period before lower mode numbers
become appreciably populated. In contrast, the $m=1$ case
exhibits an earlier redistribution of spectral energy toward
additional modes. This spectral broadening reflects nonlinear
mode coupling during the evolution of the current sheet and is
consistent with the transition from the quasi-linear to the
nonlinear regime. The rapid nonlinear acceleration itself is
identified by the reconnection-rate evolution shown in
Fig.~\ref{explrecon}.

The spectral evolution provides additional evidence that the
rapid nonlinear acceleration identified in
Fig.~\ref{explrecon} is associated with nonlinear dynamics.
Before the onset of rapid reconnection, the spectrum remains
comparatively localized around the initially excited
perturbation and its weak harmonics. After the onset, however,
a broad range of modes is rapidly excited, indicating nonlinear
mode coupling and energy transfer across Fourier modes. From a
spectral perspective, this transfer toward lower mode numbers
is consistent with nonlinear beating and sideband generation,
while the lower-$m$ modes themselves possess a larger tearing
drive and therefore grow more readily. Such spectral
redistribution is not expected from the linear evolution of a
fixed operator, including transient amplification associated
with non-normality alone. We therefore interpret the rapid
nonlinear acceleration observed in Fig.~\ref{explrecon} as a
nonlinear reconnection event, which may nevertheless be
facilitated by the non-normal transient amplification discussed
above.

\begin{figure}[htp]
\centering
  \includegraphics[trim=0 0 0 0,clip,width=0.4\textwidth]{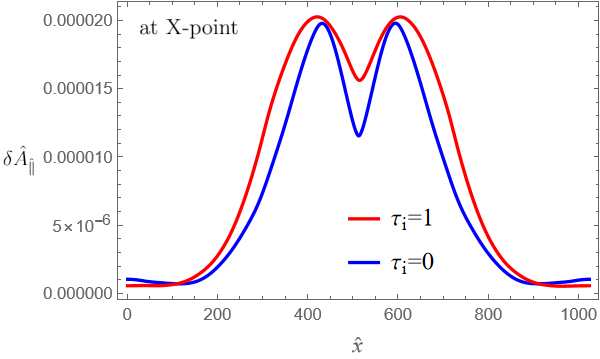}
  \includegraphics[trim=0 0 0 0,clip,width=0.4\textwidth]{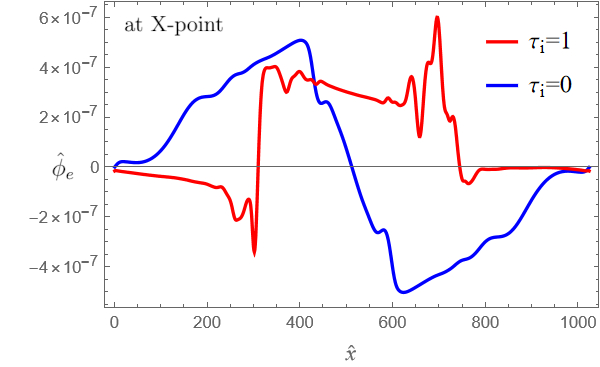}
  \includegraphics[trim=0 0 0 0,clip,width=0.4\textwidth]{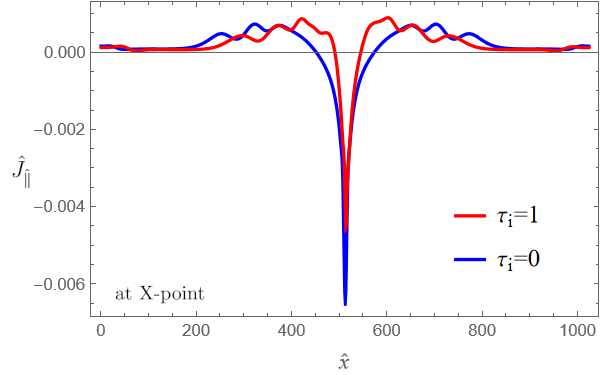}
  \includegraphics[trim=0 0 0 0,clip,width=0.4\textwidth]{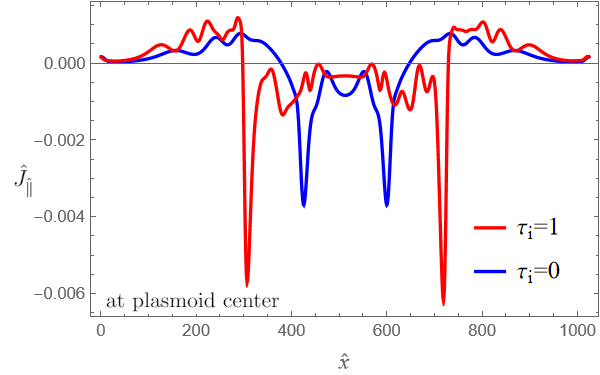}
  \caption{1D slices of fields in $x$-direction at the beginning of the magnetic reconnection, when the X-point is formed. Although the observed reconnection rates are often similar, the time evolution of the exact quantities differs, including an asymmetry in the potential $\hat {\phi}_e$ and a weaker asymmetry in $\hat{J}_{\hat{\parallel}}$ at the X-point and plasmoid center. }
  \label{cuts}
\end{figure}

To investigate the time evolution of the current sheet and possible FLR effects, $x$-cuts of several quantities are plotted and compared at different times with results presented in the literature~\cite{ScottV2,Grasso}. Figure~\ref{cuts} shows slices of $\delta \hat{A}_{\hat{\parallel}}$ at the time when the $x$-point emerges, and changes in the indent of $\hat{A}_{\hat{\parallel}}$ due to ion FLR effects are observed, similar to those reported in~Ref. \citenum{Grasso}, together with narrowing of the current sheet and island formation. However, at this stage of the reconnection, a larger island width is observed when ion temperature effects are included, and the electric potential is found to display the expected multipolar structure. The characteristic left/right asymmetry associated with ion
FLR effects is clearly visible in the electrostatic potential
$\hat{\phi}_e$,
while only a comparatively weak asymmetry is observed in
the parallel current density
$\hat{J}_{\hat{\parallel}}$.

When the reconnection is almost completed and possible elongation of the reconnection zone is fully developed, the indent in $\hat{A}_{\hat{\parallel}}$ has nearly vanished as constant behavior is approached. The electric potential then becomes rather chaotic with increased fine structure in the warm ion $\tau_i=1$ case, which translates via $N_z$ to more fine structure in the current $\hat{J}_{\hat{\parallel}}$.

\section{Plasmoid Formation and Aspect ratio effects}\label{plamoresu}
\begin{figure}[htp]
\centering

\begin{minipage}{0.05\textwidth}
\raggedright \textbf{a.)}
\end{minipage}
\begin{minipage}{0.40\textwidth}
\includegraphics[trim=2 0 10 0,clip,width=\textwidth]{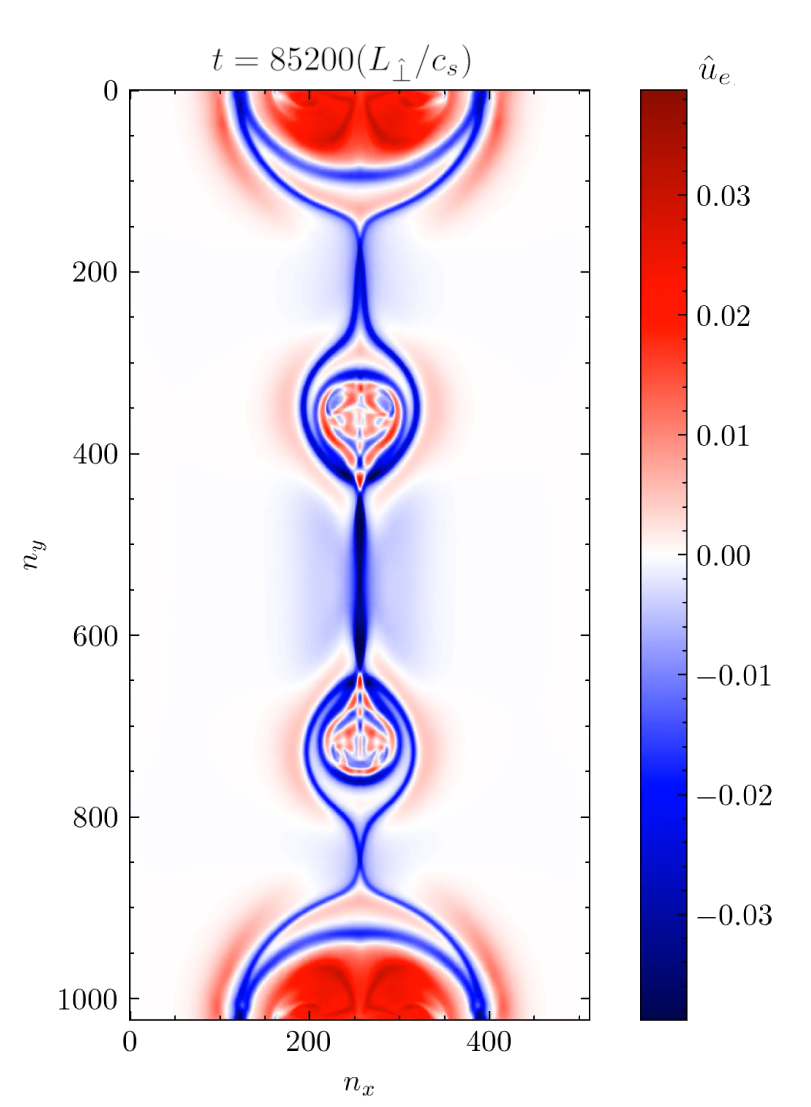}
\end{minipage}
\hspace{0.05\textwidth}
\begin{minipage}{0.05\textwidth}
\raggedright \textbf{b.)}
\end{minipage}
\begin{minipage}{0.40\textwidth}
\includegraphics[trim=2 0 10 0,clip,width=\textwidth]{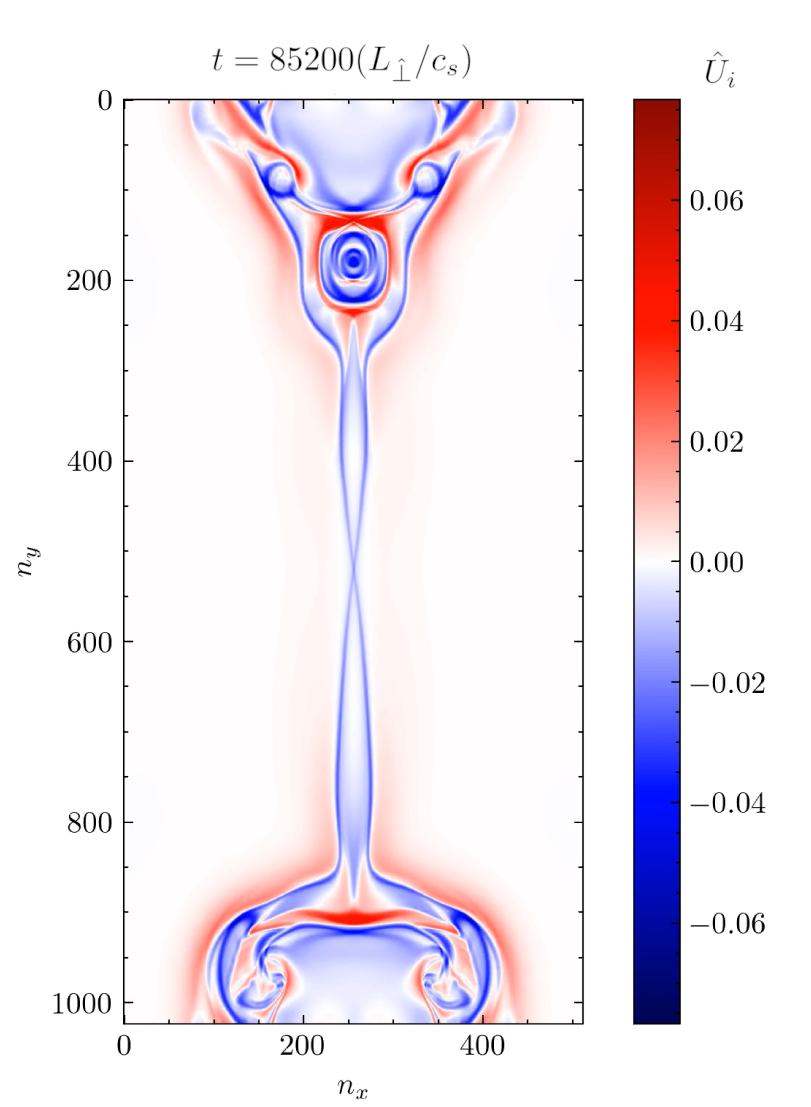}
\end{minipage}

\caption{Full-$F$ parallel species-velocities $\hat{U}_z (\hat{\vbm{x}})$ for an ion-current induced magnetic reconnection event with aspect ratio $\frac{L_{\hat{y}}}{L_{\hat{x}}}=2$ and $\tau_i=1$. The electron current shows two plasmoids growing, while there are no plasmoids developed in the ion current at this time of the reconnection.}
\label{plasmo}
\end{figure}

To further test the Full-$F$ Full-k framework of the code, a situation is constructed in which the initial current $\hat{J}$ is mainly carried by the ions,
\begin{equation}
-\frac{1}{\beta} \hat{\nabla }^2_\perp \hat{A}_{{\hat{\parallel}}}=\hat{\Gamma}_1(\hat{U}_i),
\end{equation}
as shown in Fig.~\ref{plasmo}, where the species velocities $\hat{U}_z$ are shown seperately. In this regime, not only a disparity in the current evolution due to $\mu_e$ is expected, but the ion contribution in the polarization equation also becomes relevant when the gyrocenter density amplitudes deviate significantly from the initial (or mean) value $N_0$. Owing to the low electron inertia, a significant electron current is formed as a response, thereby contributing to the reconnection process. The resulting configuration exhibits plasmoid formation with intermediate $x$-points, which contribute to the total growth rate. Nevertheless, the obtained growth rate is found to be hardly comparable with the case in which the electron current is used for the initialisation, and the growth rates are generally smaller by a factor of $10$ for all values of $\beta$. We further point out that the plasmoid formation appears to happen almost independently in each species velocity $\hat{U}_z$. This might be a result of the initial condition, and any real situation with a coherent history might never show such behaviour.

In the regime where a significant fraction of the equilibrium
current is carried by the ions, ion-FLR effects are expected
to influence the nonlinear reconnection dynamics. We therefore
consider an ion-current-supported Harris-type sheet and scan
the ion temperature ratio $\tau_i$ and electron skin depth
$\hat d_e=d_e/\rho_s$ within the Full-$F$ formulation.

For a square domain with $\beta=10^{-4}$, electron mass ratio
$\mu_e=-1/1836$, and warm ions $\tau_i=1$, the reconnection
develops the characteristic quadrupolar electrostatic potential
structure around the X-point, which has previously been
associated with stabilization and elongation of the
reconnection zone~\cite{Liu}. Although this behavior is also
observed in our simulations, it is not shown explicitly in the
present manuscript.

\begin{center}
\begin{figure*}[htp]
\centering
\includegraphics[trim=0 0 0 0,clip,width=1.0\textwidth]
{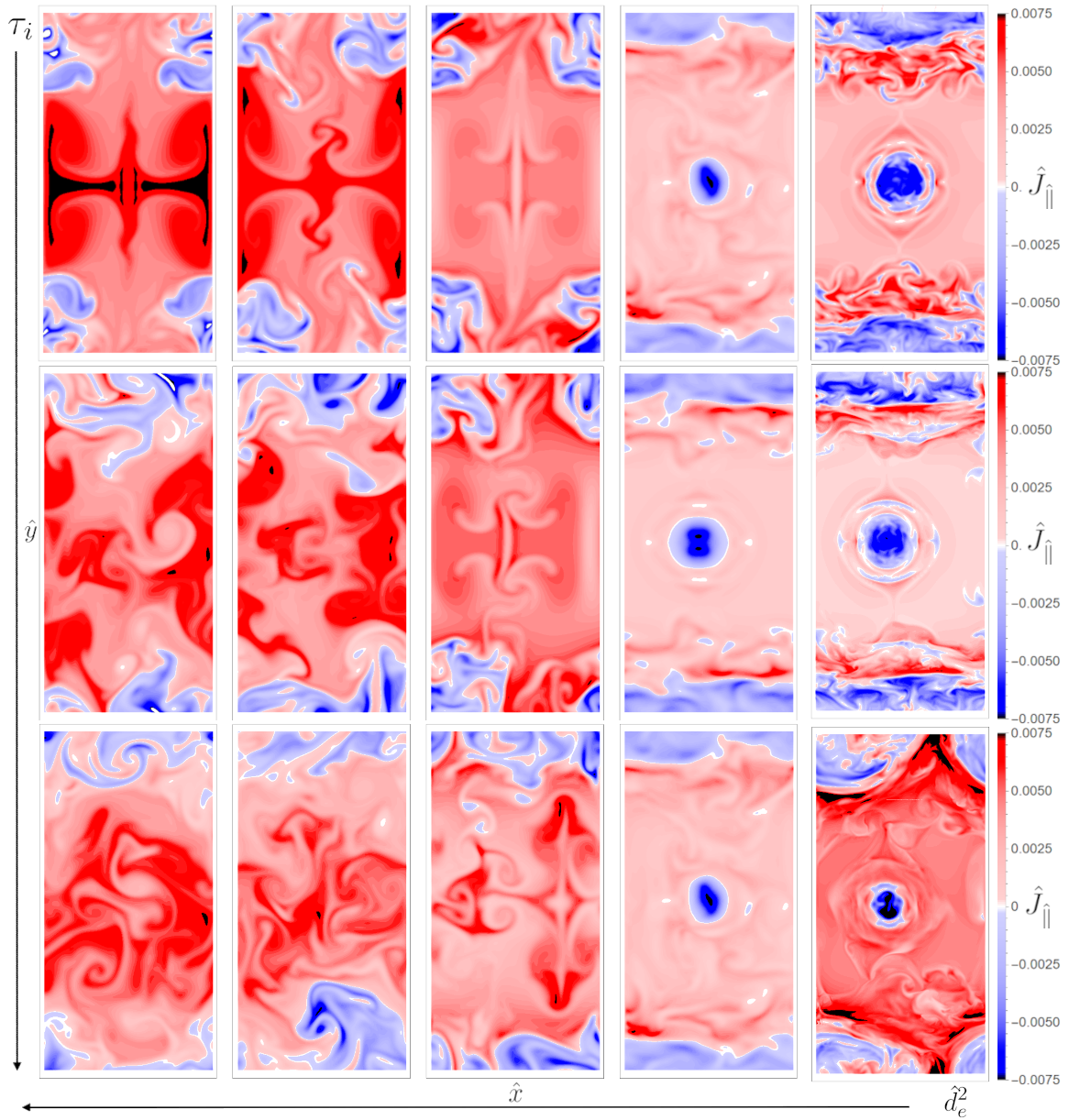}
\caption{
Snapshots of the parallel current density for an
ion-current-supported sheet with aspect ratio
$L_{\hat y}/L_{\hat x}=2$ in the Full-$F$ formulation.
The columns correspond to ion temperature ratios
$\tau_i \in \{0,1,5\}$, while the rows correspond to
$\hat d_e \in
\{5446.62,\,544.662,\,54.4662,\,5.44662,\,0.544662\}$.
Each snapshot is taken at a comparable nonlinear stage, when
the dominant magnetic island has formed. Finite ion temperature
modifies the local current-sheet morphology, but no systematic
increase in small-scale structure or in the number of dominant
plasmoids is observed with increasing $\tau_i$. A single main
island is present in all cases shown here.
}
\label{betacomp1}
\end{figure*}
\end{center}

\begin{center}
\begin{figure*}[htp]
\centering
\includegraphics[trim=0 0 0 0,clip,width=1.0\textwidth]
{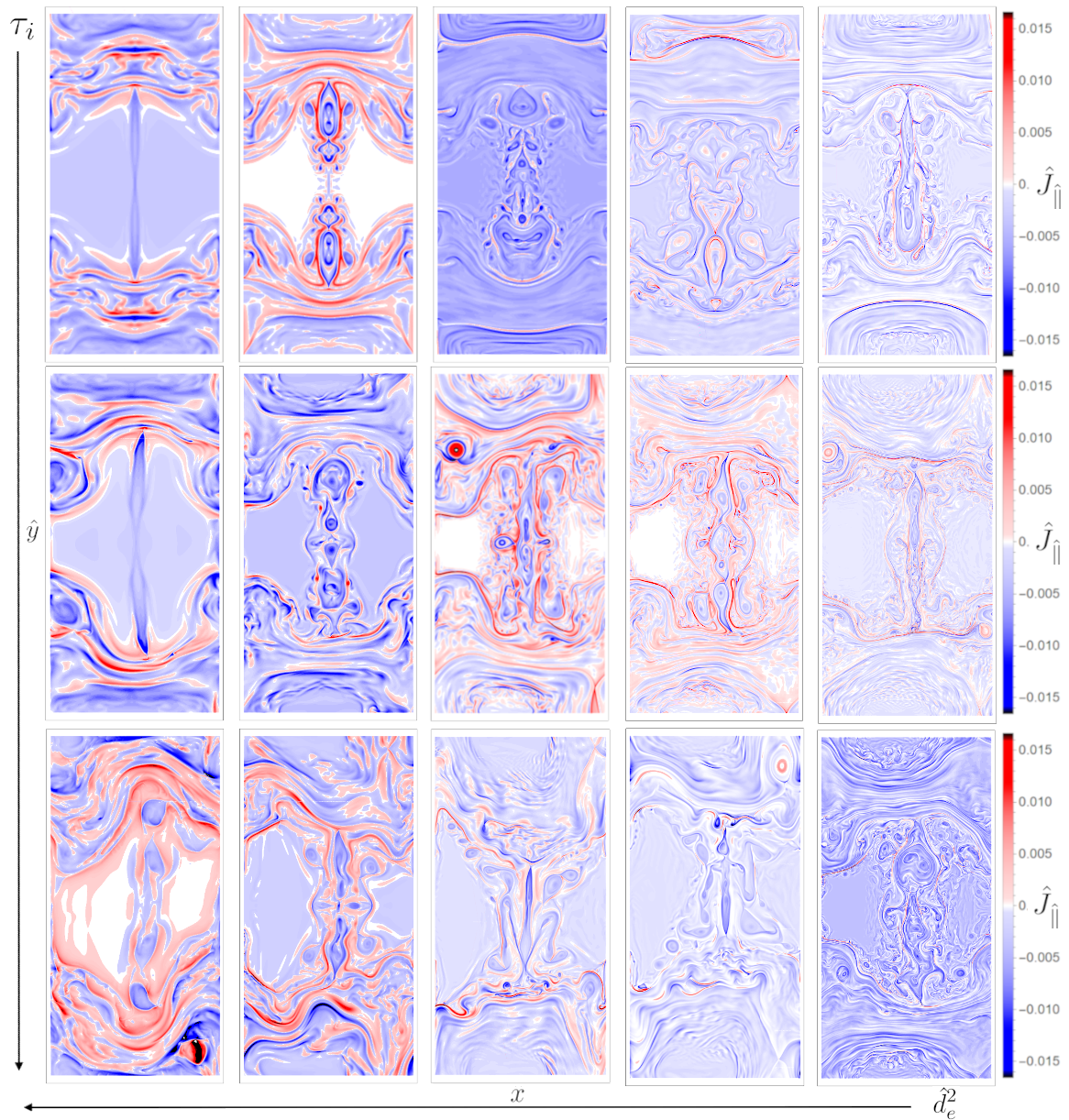}
\caption{
Snapshots of the parallel current density for an
ion-current-supported sheet with aspect ratio
$L_{\hat y}/L_{\hat x}=2$ in the Full-$F$ formulation.
The columns correspond to ion temperature ratios
$\tau_i \in \{0,1,5\}$, while the rows correspond to
$\hat d_e \in
\{5.44662\times10^{-2},\,5.44662\times10^{-3},\,
5.44662\times10^{-4},\,5.44662\times10^{-5},\,
5.44662\times10^{-6}\}$.
Each snapshot is shown at a comparable nonlinear stage of the
reconnection event. For decreasing $\hat d_e$, secondary island
formation and small-scale current-sheet structure become more
pronounced. Although finite ion temperature modifies the local
current-sheet morphology, the dominant plasmoid structure is
mainly controlled by the tearing and plasmoid dynamics, and no
systematic change in the number of dominant plasmoids is observed
with increasing $\tau_i$.
}
\label{betacomp2}
\end{figure*}
\end{center}

We next consider the same Harris-type-sheet initial condition
in an elongated domain with aspect ratio
$L_{\hat y}/L_{\hat x}=2$, for which secondary island formation
is expected. Figures~\ref{betacomp1} and~\ref{betacomp2}
show Full-$F$ simulations for two consecutive ranges of
$\hat d_e$. Each panel is taken at a comparable nonlinear stage
of the corresponding reconnection event, when the dominant
magnetic island or secondary island structure has formed.

Taken together, Figs.~\ref{betacomp1} and~\ref{betacomp2}
represent a continuous qualitative scan over $\hat d_e$. The
transition from a single dominant island to a more structured
plasmoid-mediated state occurs approximately between the
smallest value shown in Fig.~\ref{betacomp1},
$\hat d_e \simeq 0.54$, and the largest value shown in
Fig.~\ref{betacomp2}, $\hat d_e \simeq 0.054$.

The comparison further shows that increasing $\tau_i$ modifies
the local current-sheet morphology, but does not produce a
systematic change in either the number of dominant plasmoids or
the amount of small-scale structure across the investigated
parameter range. Instead, the transition from a single dominant
island to a more structured plasmoid-mediated state is mainly
controlled by $\hat d_e$ and by the nonlinear competition
between tearing modes. We emphasize that the statements
regarding the number of plasmoids and the separation into two
$\hat d_e$ regimes are qualitative observations from the
present parameter scan rather than quantitative scaling laws.
Thus, in the simulated case and parameter range, ion-FLR effects
influence the detailed morphology of the reconnecting layer,
while the dominant plasmoid topology is primarily set by the
aspect ratio, electron skin depth, and tearing-mode dynamics.

%\subsection{Aspect Ratio Effects}\label{aspeff}

\begin{center}
\begin{figure*}[]
\centering
\includegraphics[trim=0 0 0 0,clip,width=1.0\textwidth]
{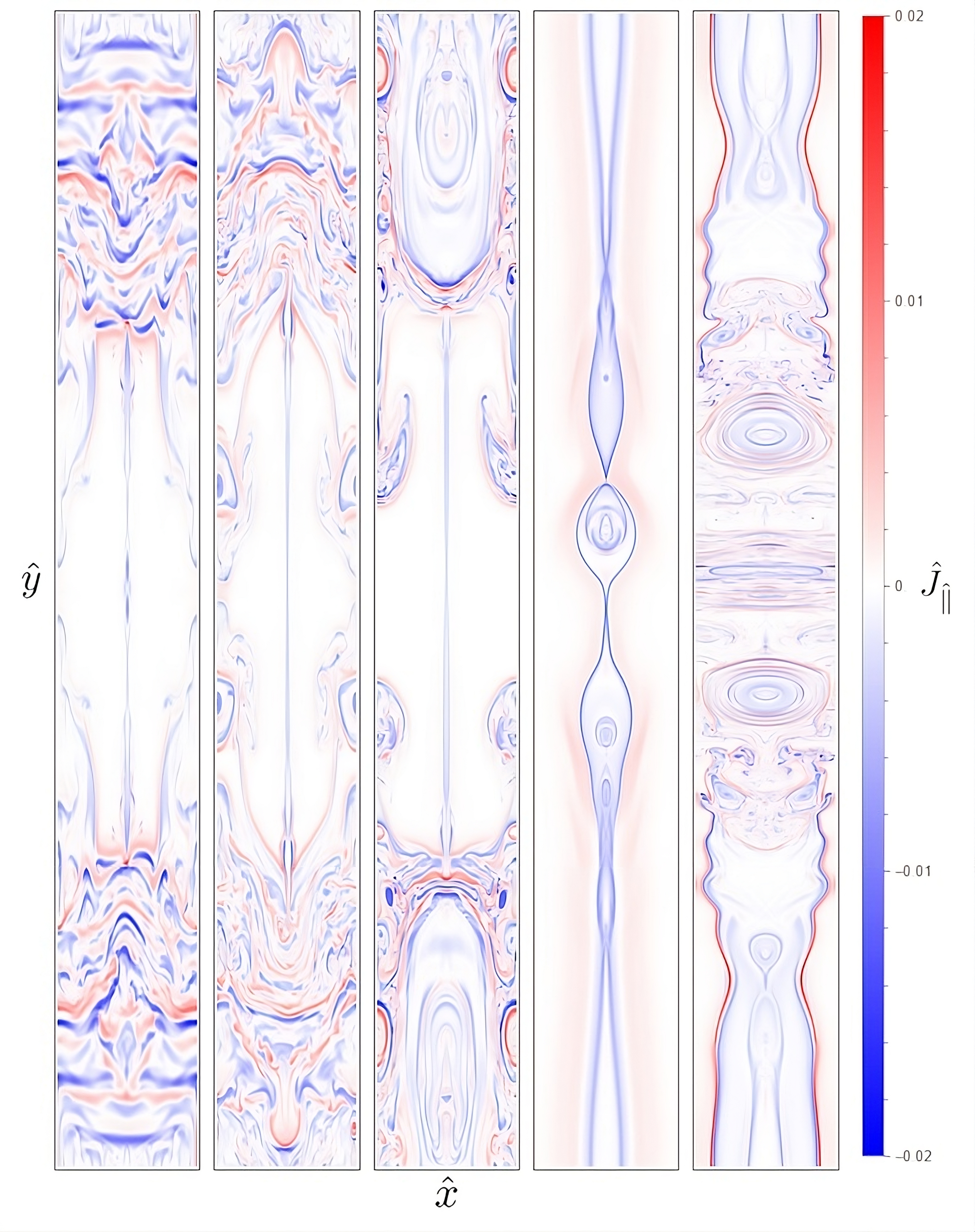}
\caption{
Full-$F$ simulations of an ion-current-supported Harris sheet
with aspect ratios
$r\in\{2,3,4,8,16\}$
from left to right, using $\tau_i=1$ and LWL polarization.
The number and characteristic size of the plasmoids generally
increase with aspect ratio. At the highest aspect ratios,
large islands are accompanied by regions containing multiple
smaller islands.
}
\label{aspectcomp}
\end{figure*}
\end{center}

To investigate the general behavior with respect to the aspect
ratio, a Full-$F$ system perturbed with $m=1$ and
$\tau_i=1$ is considered, and aspect ratios of up to~$L_{\hat{y}}/L_{\hat{x}}=16$  are
examined. Figure~\ref{aspectcomp} shows that the number of
plasmoids generally increases with increasing aspect ratio.
From an aspect ratio of~$8$ onwards, reconnection is observed to
proceed through large plasmoids whose sizes are comparable to
those of the main islands. In these configurations, the phases
of reconnection vary strongly among the multiple X-points, such
that reconnection may already be completed at some locations
while still being at an early stage at others. At the highest
aspect ratios considered, multiple regions containing small
islands are observed around the main islands.

\section{Conclusion and Summary}\label{Concl}

We have investigated collisionless magnetic reconnection and plasmoid formation in a two-dimensional Full-$F$ gyrofluid framework designed for fusion-relevant, strongly magnetized, low-$\beta$ regimes. Using the recently developed \textsc{GREENY} code~\cite{Locker3}, we quantified tearing-mode growth, transient non-modal amplification, and nonlinear explosive reconnection, and mapped the onset and morphology of plasmoid dynamics as functions of electron inertia, ion FLR effects, polarization modeling, and system aspect ratio.

A main outcome is that the Full-$F$ gyrofluid model reproduces key signatures of fast collisionless reconnection reported in Hall-MHD, gyrokinetic, and kinetic simulations, with peak normalized reconnection rates of order $E_{\mathrm{rec}}/(B_0 v_A)\sim 0.1$--$0.2$ in the parameter range studied~\cite{Shay1999,Birn2001,Yamada2010,Granier3}. In the marginally stable regime, numerically measured linear growth rates agree well with analytical tearing-mode estimates when arbitrary-wavelength polarization is retained, whereas long-wavelength and Oberbeck--Boussinesq polarization approximations can introduce noticeable deviations at small $\beta$ and intermediate $\hat d_e$. This demonstrates that consistent polarization physics is essential for quantitative reconnection studies at $\rho_s$ scales.

The linearized Full-$F$ system is shown to be strongly non-normal. Large condition numbers, a positive numerical abscissa, and $\varepsilon$-pseudospectra extending into the unstable half-plane indicate that transient amplification is intrinsic to the linearized dynamics. This non-modal amplification should not be interpreted as a direct description of the nonlinear explosive phase itself. Rather, it shows that perturbations may be amplified efficiently before or during the transition into the nonlinear regime, beyond what is predicted by asymptotic modal growth alone. The explosive reconnection phase observed in the nonlinear simulations is instead associated with mode coupling, spectral broadening, and topological rearrangement of the reconnecting magnetic field. Thus, the non-modal analysis provides a possible pathway for enhancing perturbation amplitudes, while the explosive phase itself remains a nonlinear reconnection process.

Ion FLR effects are found to be secondary for linear growth rates in square domains, but they influence current-sheet morphology, island width, and the fine structure of the reconnection region in warm-ion regimes, consistent with earlier gyrofluid and reduced-kinetic studies~\cite{Grasso,Biancalani}. In particular, when a significant fraction of the equilibrium current is carried by ions, FLR effects contribute to the shaping of the reconnection geometry and modify the local structure of plasmoid-mediated reconnection in elongated systems through enhanced species disparity and nonlinear coupling.

The aspect-ratio comparison shows that plasmoid formation is promoted in elongated current sheets. With increasing $L_{\hat{y}}/L_{\hat{x}}$, the number of plasmoids generally increases, yielding multiple X-points and magnetic islands together with increasingly spatially inhomogeneous reconnection dynamics. At the largest aspect ratios considered, reconnection proceeds through coexisting large islands and surrounding smaller island structures, indicating a transition toward a multiscale reconnection regime. These observations are qualitative and should not be interpreted as a quantitative scaling law.

Numerically, we emphasize that centered differencing requires explicit regularization. In our implementation this is achieved via a weak hyperdiffusion term, chosen small enough to avoid artificially broadening the reconnecting current sheet while still suppressing grid-scale noise. Importantly, Fig.~\ref{Upwind} demonstrates that an Arakawa discretization with $\nu=10^{-7}$ and an upwind scheme yield very similar parallel-current structures for both the arbitrary-wavelength NOB and NOB+LWL polarization models, while reconnection is consistently accelerated when arbitrary-wavelength NOB polarization is employed. This comparison supports two key conclusions: (i) the qualitative and quantitative reconnection dynamics reported here are not artifacts of the discretization choice, and (ii) the reconnection acceleration observed with arbitrary-wavelength NOB polarization reflects a physical model difference rather than numerical bias.

Finally, while convergence is robust for long-wavelength polarization at moderate resolution, arbitrary-wavelength polarization introduces $\rho_s$-scale structure whose detailed origin and sensitivity to numerical dissipation merit further investigation. Future work should therefore include systematic dissipation-parameter scans, improved iterative solvers for polarization/Amp\`ere closures, and extensions toward three-dimensional coupling, such as shear-Alfv\'enic dynamics, and reconnection in the presence of drift-wave turbulence. Overall, the present results establish Full-$F$ gyrofluid modeling as a quantitatively consistent and computationally efficient framework for studying fast reconnection, transient amplification, and plasmoid dynamics in tokamak-relevant parameter regimes.

\begin{acknowledgments}
This work has been carried out within the framework of the EUROfusion Consortium, funded by the European Union via the
Euratom Research and Training Programme (Grant Agreement No 101052200 — EUROfusion). Views and opinions
expressed are however those of the author(s) only and do not necessarily reflect those of the European Union or the European
Commission. Neither the European Union nor the European Commission can be held responsible for them.\\

This work was supported by the UiT Aurora Centre Program, UiT The Arctic University of Norway (2020). This research was funded in whole or in part by the Austrian Science Fund (FWF) [P 34241-N]. \\

 The author thanks Bruce Scott and Matthias Wiesenberger (DTU) for valuable discussions.\\
\end{acknowledgments}

\subsection{Author Contributions}
\textbf{F. F. Locker}: Data Curation (lead); Conceptualization (equal); Validation (equal); Software (equal); Visualization (equal); Writing - original draft (lead); Methodology (equal); Formal analysis (lead); Project Administration (equal);
\textbf{M. Rinner}: Validation (supporting); Visualization (equal); Software (supporting);
\textbf{M. Held}: Conceptualization (equal); Methodology (equal);
\textbf{A. Kendl}: Supervision (lead); Conceptualization (equal); Writing – review \& editing (equal); Resources (lead); Project Administration (equal);
\section*{Data Availability Statement}
The full code is available on \url{https://git.uibk.ac.at/c7441315/greeny} and the presented data that supports the findings of this study are available from the corresponding author upon reasonable request.

\appendix

\section{Derivation of the Dispersion Relation}\label{AppendixA}

After neglecting second order terms, we find the linearised system of equations 

\begin{align*}
    %%%%%%%%%%%%%%%%%%%%%%%%%%%%%%%%%%%% 1
    %1] \hspace{2cm}
    \mathrm{i}\frac{g}{\beta}\,\tilde{n}_e 
    =  - a_{||,\mathrm{eq}}'\,
     &\left(\frac{A_0}{N_0}\,a_{||,\mathrm{eq}}''\,\tilde{n}_e
    + N_0\,\tilde{U}_e\right)
    +\\
    +& \,a_{||,\mathrm{eq}}'''\,\tilde{A}_{||},
    %\tag{1$^\mathrm{L3}$}\label{eq:4.2_LinStep3_1}
    \tag{1$^\mathrm{L}$}\label{eq:4.2_Linearized_eq1}
    \\
    %
    %
    %%%%%%%%%%%%%%%%%%%%%%%%%%%%%%%%%%%% 2
    %2] \hspace{2cm}
    \mathrm{i}\frac{g}{\beta}\,\tilde{N}_i
    = & - a_{||,\mathrm{eq}}'\,N_0\,\tilde{U}_i,
    %\tag{2$^\mathrm{L3}$}\label{eq:4.2_LinStep3_2}
    \tag{2$^\mathrm{L}$}\label{eq:4.2_Linearized_eq2}
    \\
    %
    %
    %%%%%%%%%%%%%%%%%%%%%%%%%%%%%%%%%%%% 3
    %3] \hspace{2cm}
    i \frac{g}{\beta}\,\left(\tilde{U}_e + \frac{1}{\mu_e} \, \beta\tilde{A}_{||}\right) &=
    - \left( \frac{1}{\beta N_0}\,a_{||,\mathrm{eq}}''' + \frac{1}{\mu_e}\,a_{||,\mathrm{eq}}' \right)\,\tilde{\phi}+\\
    +\frac{1}{\mu_e N_0}\,a_{||,\mathrm{eq}}'\,\tilde{n}_e-\frac{A_0}{  N_0} \,a_{||,\mathrm{eq}}'' &\,\left ( a_{||,\mathrm{eq}}'\,\tilde{U}_e - \frac{1}{ N_0}\,a_{||,\mathrm{eq}}'''\,\tilde{A}_{||}\right),
    %\tag{3$^\mathrm{L3}$}\label{eq:4.2_LinStep3_3}
    \tag{3$^\mathrm{L}$}\label{eq:4.2_Linearized_eq3}
    \\
    %
    %
    %%%%%%%%%%%%%%%%%%%%%%%%%%%%%%%%%%%% 4
    %4] \hspace{2cm}
   \frac{\mathrm{i}}{\beta}g\,\left(\tilde{U}_i + \beta\tilde{A}_{||}\right) = &
    -a_{||,\mathrm{eq}}'\,\tilde{\phi},
    %\tag{4$^\mathrm{L3}$}\label{eq:4.2_LinStep3_4}
    \tag{4$^\mathrm{L}$}\label{eq:4.2_Linearized_eq4}
    \\
    %
    %
    %%%%%%%%%%%%%%%%%%%%%%%%%%%%%%%%%%%% 5
    %5] \hspace{2cm}
    -\,\left(\tilde{A}_{||}''-k^2 \tilde{A}_{||}\right)
    = &\,\, N_0\,\left(\tilde{U}_i - \tilde{U}_e\right)
    - \frac{A_0}{ N_0}\,a_{||,\mathrm{eq}}''\,\tilde{n}_e,
    %\tag{5$^\mathrm{L3}$}\label{eq:4.2_LinStep3_5}
    \tag{5$^\mathrm{L}$}\label{eq:4.2_Linearized_eq5}
    \\
    %
    %
    %%%%%%%%%%%%%%%%%%%%%%%%%%%%%%%%%%%% 6
    %6] \hspace{2cm}
    -\left(\tilde{\phi}'' - k^2 \tilde{\phi}\right)
    = &\,\, \frac{1}{N_0}\,\left(\tilde{N}_i - \tilde{n}_e\right).
    %\tag{6$^\mathrm{L3}$}\label{eq:4.2_LinStep3_6}
    \tag{6$^\mathrm{L}$}\label{eq:4.2_Linearized_eq6}
\end{align*}
where we introduced the quantity $g := \frac{\gamma}{k\, A_0}$. From this set one can derive another equation by subtracting equation 
(\ref{eq:4.2_Linearized_eq1})
from  
(\ref{eq:4.2_Linearized_eq2})
and substituting the $\left(\tilde{U}_i - \tilde{U}_e\right)$ term with equation 
(\ref{eq:4.2_Linearized_eq5})
and the $\left(\tilde{N}_i - \tilde{n}_e\right)$ term with equation 
(\ref{eq:4.2_Linearized_eq6}).
Conducting this, one gets
\begin{align*}
    %
    %
    %%%%%%%%%%%%%%%%%%%%%%%%%%%%%%%%%%%% 7
    \mathrm{i} g\, N_0\,(\tilde{\phi}'' -k^2 \tilde{\phi})
    -  a_{||,\mathrm{eq}}'''\,\beta\tilde{A}_{||}
    + a_{||,\mathrm{eq}}'\, (\beta\tilde{A}_{||}'' -k^2 \beta\tilde{A}_{||}) = 0.
    \tag{7$^\mathrm{L}$}\label{eq:4.2_Linearized_eq7}
\end{align*}
 This last equation is very fundamental and found as an intermediate result in many other derivations of the tearing instability parameter \cite{Fitzpatrick,Tassi}. 
 
 \subsection{Solving for different Regimes}
Solving this system of equations for $\tilde{A}_{\hat{\parallel}}$ and $\tilde{\phi}$ we use the ordering

\begin{align}
\hat{\Delta} ' d_e \ll 1, \quad \gamma \ll 1, \quad d_e^2 \ll 1,
\end{align}
to obtain an analytically solvable differential equation. Substitutions, rearrangements and simplifications were done by pen and paper and double checked with the CAS Wolfram Mathematica.\cite{Mathematica} For small growth rates $\gamma$ we can decouple $\phi$ and $\tilde{A}_{\hat{\parallel}}$ in equation \ref{eq:4.2_Linearized_eq7} and obtain
\begin{align}
    -  a_{||,\mathrm{eq}}'''\tilde{A}_{||}
    + a_{||,\mathrm{eq}}'\, (\tilde{A}_{||}'' -k^2 \tilde{A}_{||}) = 0.
\end{align}

which is a Legendre differential equation. Note that neglecting $\gamma$ terms holds only because $g$ becomes independent of $\beta$ beforehand. Otherwise that would not universally hold as $\beta \ll 1$ too. With
\begin{align*}
    a_{||,\mathrm{eq}}' &= -\tanh(x)\\
    \mathrm{and} \hspace{2cm}
    a_{||,\mathrm{eq}}''' &= 2\,\frac{\tanh(x)}{\cosh^2(x)}
\end{align*}

The solution is
\begin{align}
    \tilde{A}_{||,\mathrm{out}} &= \tilde{A}_{||,0}\,
    \begin{cases}
        \mathrm{e}^{-kx}\,\left(1+\tanh(x)/k\right), &x>0
        \\
        \mathrm{e}^{kx}\,\left(1-\tanh(x)/k\right), &x<0
    \end{cases},
    \label{eq:4.3.1_solution_A_outer_region}
\end{align}
where $\tilde{A}_{||,0}\in\mathbb{R}$ is an arbitrary constant.

The function has a discontinuous derivative at $x=0$, just as it is needed for the tearing stability parameter $\hat{\Delta}'$, which is defined by the solution of this outermost region 
\begin{align}
    \hat{\Delta}'_{outer} &= \lim_{x\to0^+}\frac{\tilde{A}_{||,\mathrm{out}}'}{\tilde{A}_{||,\mathrm{out}}}
    -\lim_{x\to0^-}\frac{\tilde{A}_{||,\mathrm{out}}'}{\tilde{A}_{||,\mathrm{out}}}
    = 2\left(\frac{1}{k}-k\right).
    \label{eq:4.3.1_delta'_parameter}
\end{align}
For small $x$ one can linearize the above solution. Near the resonant surface at $x=0$ the outer solution behaves like
\begin{align*}
    \tilde{A}_{||,\mathrm{out}}(x) \approx 
    \tilde{A}_{||,0}\,\left(1+\frac{\hat{\Delta}'}{2}\,|x|\right).
\end{align*}
Lets now solve the equations for the so called "inner region" close to the current sheet, so for $|x| \ll 1$.

By Taylor-expanding the used equilibrium function  $a_{||,\mathrm{eq}}(x) = -\ln(\cosh(x))$ one finds
\begin{align}
    a_{||,\mathrm{eq}}(x) &\approx 0,
    \hspace{2cm}
    a_{||,\mathrm{eq}}'(x) \approx -x,\\
    a_{||,\mathrm{eq}}''(x) &\approx -1,
    \hspace{1.7cm}
    a_{||,\mathrm{eq}}'''(x) \approx 2x.
\end{align}

Since all $k^2$ terms originate from a $y$ derivative, ($k=\hat{k}_{\hat{y}}$), and the variations of the functions are larger in $x$-direction, the following can be applied
\begin{align}
    k^2\tilde{\phi}\ll \tilde{\phi}'',\\
    k^2\tilde{A}_{||}\ll \tilde{A}_{||}'',\\
    \left(k^2-2\right)\tilde{A}_{||}\ll \tilde{A}_{||}'',
\end{align}

With these approximations the equations (\ref{eq:4.2_Linearized_eq1}) to (\ref{eq:4.2_Linearized_eq7}) become
\begin{align}
    %%%%%%%%%%%%%%%%%%%%%%%%%%%%%%%%%%%% 1
    \left(\mathrm{i}\frac{g}{\beta}
    +\frac{A_0}{N_0}\,x
    \right)\,\tilde{n}_e 
    = &\,\,  N_0\,x\,\tilde{U}_e
    +\,2x\,\tilde{A}_{||},
    \tag{1$^x$}\label{eq:4.3.2_smalL_eq1}
    \\
    %
    %
    %%%%%%%%%%%%%%%%%%%%%%%%%%%%%%%%%%%% 2
    \mathrm{i}\frac{g}{\beta}\,\tilde{N}_i
    = &\,\,  N_0\,x\,\tilde{U}_i,
    \tag{2$^x$}\label{eq:4.3.2_smalL_eq2}
    \\
    %
    %
    %%%%%%%%%%%%%%%%%%%%%%%%%%%%%%%%%%%% 3
    \left(\mathrm{i}\frac{g}{\beta}
    +\frac{A_0}{ N_0}\,x
    \right)\,\tilde{U}_e =
    &-\left(\mathrm{i}g\,\frac{1}{\mu_e}
    +\frac{A_0}{ N_0}\,\frac{2}{ N_0}\,x
    \right)\,\tilde{A}_{||}
    \\
    &- x\,\left( \frac{2}{\beta N_0} - \frac{1}{\mu_e} \right)\,\tilde{\phi}
    -\frac{1}{\mu_e N_0}\,x\,\tilde{n}_e,
    \tag{3$^x$}\label{eq:4.3.2_smalL_eq3}
    \\
    %
    %
    %%%%%%%%%%%%%%%%%%%%%%%%%%%%%%%%%%%% 4
    \mathrm{i}\frac{g}{\beta}\,\tilde{U}_i = & - \mathrm{i}g\,\tilde{A}_{||}
    +x\,\tilde{\phi},
    \tag{4$^x$}\label{eq:4.3.2_smalL_eq4}
    \\
    %
    %
    %%%%%%%%%%%%%%%%%%%%%%%%%%%%%%%%%%%% 5
    -\tilde{A}_{||}''
    = &\,\, N_0\,\left(\tilde{U}_i - \tilde{U}_e\right)
    +\frac{A_0}{N_0}\,\tilde{n}_e,
    \tag{5$^x$}\label{eq:4.3.2_smalL_eq5}
    \\
    %
    %
    %%%%%%%%%%%%%%%%%%%%%%%%%%%%%%%%%%%% 6
    -\tilde{\phi}''
    = &\,\, \frac{1}{N_0}\,\left(\tilde{N}_i - \tilde{n}_e\right),
    \tag{6$^x$}\label{eq:4.3.2_smalL_eq6}
    \\
    %
    %
    %%%%%%%%%%%%%%%%%%%%%%%%%%%%%%%%%%%% 7
    0 = &\,\,   g\,\tilde{\phi}''
    +\mathrm{i}\frac{x}{ N_0}\,\beta \tilde{A}_{||}''.
    \tag{7$^x$}\label{eq:4.3.2_smalL_eq7}
\end{align}
We now want to find an ODE for $\tilde{A}_{\hat{\parallel}} $. For that we start with \ref{eq:4.3.2_smalL_eq7} and eliminate $\tilde{\phi}''$ to obtain
\begin{align}
g(n_e-N_i)=\beta i x \tilde{A}_{\hat{\parallel}}''.
\end{align}
Then one has to insert for $n_e$ (eliminate $U_e$ in the electron density equation first), $U_i$ and $N_i$ to get
\begin{align}
\tilde{A}_{\hat{\parallel}}''=-\tilde{A}_{\hat{\parallel}} \beta N_0 -\frac{i\beta N_0 \phi x}{g}+\frac{2 \tilde{A}_{\hat{\parallel}} g^2 \mu_e N_0^2}{\Omega}-\\ -
\frac{\tilde{A}_{\hat{\parallel}} \beta^2 g^2 N_0^3}{\Omega}+\frac{2 i g \mu_e N_0^2 \phi x}{\Omega}-\frac{i \beta g N_0^3 \phi x}{\Omega}
\end{align}
where $\Omega= -g^2\mu_e N_0^2+2i A_0 \beta g \mu_e N_0 x +\beta^2(A_0^2 \mu_e + N_0^2)x^2$ is used to shorten it. We can rearrange that to
\begin{align}
\tilde{A}_{\hat{\parallel}}''+N_0\frac{[\Omega \beta +g^2N_0(\beta N_0-2\mu_e)](\tilde{A}_{\hat{\parallel}} g + i\phi x)}{g \Omega}=0.
\end{align}
The imaginary part of $\Omega$, $Im[\Omega]= 2A_0\beta g \mu_e N_0 x$ is relatively small and can be neglected. This can be cast into the form of
\begin{align}
\label{aparderv}
\tilde{A}_{\hat{\parallel}}''= \kappa \bigg(\tilde{A}_{\hat{\parallel}} +\frac{i\tilde{\phi}x}{g} \bigg ) 
\end{align}
where $\kappa$ is now
\begin{align}
 \kappa=\frac{N_0(g^2N_0(\beta N_0 -\mu_e(2+\beta N_0))+\beta^3(A_0^2 \mu_e + N_0^2)x^2)}{-g^2 \mu_e N_0^2  +\beta^2(A_0^2 \mu_e + N_0^2)x^2}.
\end{align}
The small $\hat{\Delta}'$ hypothesis allows one to set $\tilde{A}_{||,\mathrm{in}}$ approximately constant around the resonant surface $x=0$ (constant-$\tilde{A}_{||,\mathrm{in}}$ approximation), thus to apply this one sets $\tilde{A}_{||,\mathrm{in}}(x) = \tilde{A}_{||,\mathrm{in},0} + \tilde{A}_{||,\mathrm{in},1}(x)$ with $|\tilde{A}_{||,\mathrm{in},1}|\ll \tilde{A}_{||,\mathrm{in},0}$. Inserting this into relation \ref{aparderv} yields
\begin{align}
    \frac{1}{\kappa}\,\left(\tilde{A}_{||,\mathrm{in},0} + \tilde{A}_{||,\mathrm{in},1}\right)''
    &= \left(\tilde{A}_{||,\mathrm{in},0} + \tilde{A}_{||,\mathrm{in},1}\right) + \mathrm{i}\frac{x}{g}\,\tilde{\phi} \\
    \Leftrightarrow\hspace{1cm}\\
    \frac{1}{\kappa}\, \tilde{A}_{||,\mathrm{in},1}''
    &= \tilde{A}_{||,\mathrm{in},0}\left(1 + \frac{\tilde{A}_{||,\mathrm{in},1}}{\tilde{A}_{||,\mathrm{in},0}}\right) + \mathrm{i}\frac{x}{g}\,\tilde{\phi} \\
    \Rightarrow\hspace{1cm}
    \frac{1}{\kappa}\, \tilde{A}_{||,\mathrm{in},1}''
    &= \tilde{A}_{||,\mathrm{in},0} + \mathrm{i}\frac{x}{g}\,\tilde{\phi} \\
    \Leftrightarrow\hspace{1cm}
    \tilde{A}_{||,\mathrm{in},1}''
    &= \kappa\left(\tilde{A}_{||,\mathrm{in},0} - x\,\left(-\mathrm{i}\frac{\tilde{\phi}}{g}\right)\right).
\end{align}
Without loss of generality we can set $\tilde{A}_{||,\mathrm{in},0}=1$ and obtain
\begin{align}
    \tilde{A}_{||,\mathrm{in},1}''
    &= \kappa \left(1 - x\,\tilde{\xi}\right) 
\end{align}
where we defined the displacement function $\tilde{\xi}=-i\tilde{\phi}/g$ \cite{Tassi}.
At this point it is helpful to rewrite $\kappa$ and defining the new quantities

\begin{align}
	\varphi^{-1}:=N_0^2(1-\mu_e(1+\frac{2}{\beta N_0}))\\
	\lambda=\beta^2(A_0^2\mu_e + N_0^2)
 \end{align}
 to obtain
 \begin{align}
     \kappa &=\frac{N_0(g^2N_0(\beta N_0 -\mu_e(2+\beta N_0))+\beta^3(A_0^2 \mu_e + N_0^2)x^2)}{-g^2 \mu_e N_0^2  +\beta^2(A_0^2 \mu_e + N_0^2)x^2}= \notag \\
   &= N_0 \beta \frac{\sigma+\frac{x^2}{\delta^2}}{\epsilon+\frac{x^2}{\delta^2}}
\end{align}

where we introduce: $\delta := \sqrt{g}$ as a renormalization factor, the parameter $\sigma := g\,\varphi^{-1}\lambda^{-1}$ and the expansion parameter $\epsilon := -g N_0^2\,(\mu_e)\lambda^{-1}$. 
We now want to first solve the ODE for $\phi$ or $\xi(x)$. So we use equation \ref{eq:4.3.2_smalL_eq7} to eliminate $\tilde{A_{\hat{\parallel}}}$

\begin{align}
\label{relAxi}
\tilde{A}_{{\hat{\parallel}},in}=N_0 \beta \frac{\sigma+\frac{x^2}{\delta^2}}{\epsilon+\frac{x^2}{\delta^2}}\left(1 - x\,\tilde{\xi}\right) = \frac{ig\tilde{\phi}''}{x\beta}N_0
\end{align}

 and after transformation of
\begin{align}
i\frac{g}{x}\tilde{\phi}''=\beta^2\frac{\sigma+\frac{x^2}{\delta^2}}{\epsilon+\frac{x^2}{\delta^2}}\bigg (1-x(-1)\frac{\tilde{\phi}}{g}  \bigg )\\
-\frac{g^2}{x}\tilde{\xi}''=\beta^2\frac{\sigma+\frac{x^2}{\delta^2}}{\epsilon+\frac{x^2}{\delta^2}}\bigg (1-x\tilde{\xi}  \bigg )
\\
\frac{g^2}{x}\tilde{\xi}''=-\beta^2\frac{\sigma+\frac{x^2}{\delta^2}}{\epsilon+\frac{x^2}{\delta^2}}\bigg (1-x\tilde{\xi}  \bigg )
\\
\frac{g^2}{x}\tilde{\xi}''=\beta^2\frac{\sigma+\frac{x^2}{\delta^2}}{\epsilon+\frac{x^2}{\delta^2}}\bigg (x\tilde{\xi} -1 \bigg )
\end{align}
with
\begin{align}
    z := \frac{x}{\delta},
    \hspace{1cm}&\hspace{1cm}
    \hat{\xi}(z) := \delta\,\tilde{\xi}(\delta z),
    \label{eq:4.3.2_renormalization}\\
    \Rightarrow\qquad
    \hat{\xi}''(z) &= \delta\,\frac{\mathrm{d}^2}{\mathrm{d} z^2}\tilde{\xi}(x)
    \nonumber\\
    &= \delta\,\frac{\mathrm{d}}{\mathrm{d} z}
    \left(\frac{\mathrm{d} x}{\mathrm{d} z}\,\frac{\mathrm{d}}{\mathrm{d} x}\tilde{\xi}(x)\right)
    \nonumber\\
    &= \delta^2\,
    \frac{\mathrm{d} x}{\mathrm{d} z}\,\frac{\mathrm{d}}{\mathrm{d} x}\tilde{\xi}'(x)
    \nonumber\\
    &= \delta^3\,\tilde{\xi}''(x),
    \nonumber
\end{align}
This equation turns into a second order differential equation in $z$ for $\hat{\xi}$
\begin{align}
     \frac{g^2}{\delta^3} \hat{\xi}''(z)=\frac{\delta^4}{\delta^3} \hat{\xi}''(z)=\beta^2 z \delta (z \hat{\xi}-1)\frac{z^2+\sigma}{z^2+\epsilon}\\
 \hat{\xi}''(z)=\beta^2 z (z \hat{\xi}-1)\frac{z^2+\sigma }{ z^2+\epsilon}.   
\end{align}
This differential equation can be solved analytically in the case where $ \epsilon \ll \sigma   \ll 1$ ordering is applied (large $z$ ordering) 
\begin{align}
 \hat{\xi}''=- \beta^2 z +  \beta^2 \hat{\xi}z^2
\end{align}
and the boundary conditions 
\begin{align}
\hat{\xi}(0)=0, \qquad \lim_{x \rightarrow \infty} \hat{\xi}(x)=x^{-1}
\end{align}
apply. The solution, given by Goldston, Rutherford and Furth, is \cite{ScottV2}
\begin{align}
\hat{\xi}(z)=\beta \frac{z}{2}\int_0^{\pi/2}\mathrm{exp}\bigg ( -\beta \frac{z^2}{2}\, \mathrm{cos}(\theta)\bigg)\mathrm{sin}^{1/2}(\theta)d\theta.
\end{align}

Having an expression for $\hat{\xi}$ we now can try to solve the integral for the matching parameter, making use of the relation for $\tilde{A}_{\hat{\parallel}}^{''}$ (eq. \ref{relAxi})
\begin{align}
\hat{\Delta} '=\int_{\infty}^\infty  \ dx  \tilde{A}_{\hat{\parallel}} ''=\\
=\int_{\infty}^\infty dx N_0 \beta \frac{\sigma+\frac{x^2}{\delta^2}}{\epsilon+\frac{x^2}{\delta^2}}\left(1 - x\,\tilde{\xi}\right)
=\\
=\int_{\infty}^\infty dz \delta N_0 \beta(1-z \hat{\xi})\frac{z^2+\sigma }{ z^2+\epsilon}
\end{align}

An expression for $\tilde{A}_{||,\mathrm{in},1}''$ was obtained by substituting via equation (\ref{eq:4.3.2_renormalization}) and rearranging to
\begin{align}
   \hat{\Delta}'_{inner} 
    &=  \beta N_0\,\int_{-\infty}^{\infty}
    \frac{(x/\delta)^2 + \sigma}{(x/\delta)^2 + \epsilon}\left(1 - \frac{x}{\delta}\,\delta\,\tilde{\xi}(x)\right)\\
    &=  \beta N_0\,\delta\,\int_{-\infty}^{\infty}
    \frac{z^2 + \sigma}{z^2 + \epsilon}\left(1 - z\,\hat{\xi}(z)\right)
    \,\mathrm{d}z\\
    &=  \beta N_0\,\delta\,\int_{-\infty}^{\infty}
    \left(
    1 + \frac{\sigma}{\epsilon}\,\frac{1}{1+z^2/\epsilon} - \frac{1}{1+z^2/\epsilon}
    \right)
    \left(1 - z\,\hat{\xi}(z)\right)
    \,\mathrm{d}z\\
    &=: \beta N_0\,\delta\,\left(
    I_1
    +\left(\frac{\sigma-\epsilon}{\epsilon})\right)\,
    I_2\right),
\end{align}
The main contribution to the integral is in the regions for $|z| > \sigma$ or $z^2 \gg \sigma \gg \epsilon$. Evaluating the first Integral one finds \cite{Furth, Bussac}
%
%
%\begin{align*}
%    I_1 
%    &= \int_{-\infty}^\infty (1-z\,\hat{\xi}(z))\,\mathrm{d}z\\
%    &\approx (\beta N_0)^{1/2}\,2.12.
%\end{align*}
\begin{align}
    I_1 
    &= \int_{-\infty}^\infty (1-z\,\hat{\xi}(z))\,\mathrm{d}z\\
    &\approx \frac{2.12}{\sqrt{\beta}}
\end{align}
which can be tested by numerical evaluation of the integral. 
Outside of the region $z\in[-\sigma,\sigma]$ the function
$$
f_2(z)=\frac{1}{1+z^2/\epsilon}
$$
 practically vanishes and suppresses every function it gets multiplied to. Thus the function $f_2(z)\,(1-z\,\hat{\xi}(z))$ is only non-zero inside this region. Together with the leading order solution valid for $|z|\lesssim \sigma/\alpha$ (has been numerically tested by us for small $\epsilon$) one concludes
\begin{align}
    I_2 
    &= \int_{-\infty}^{\infty}\frac{1 - z\,\hat{\xi}(z)}{1+z^2/\epsilon}\,\mathrm{d}z
    \nonumber\\
    &\approx \pi \epsilon^{1/2}.
    \label{eq:4.3.3_result_I_2}
\end{align}
\begin{align}
    \hat{\Delta}'_{inner} &= \beta N_0\,\delta\,\left( \frac{2.12}{\sqrt{\beta}}
    +\left(\frac{\sigma-\epsilon}{\epsilon}\right)\,
    \pi\,\epsilon^{1/2}\right).
\end{align}

Substituting the definitions one finally gets
\begin{align}\begin{split}
        \hat{\Delta}'_{outer}=2\left(\frac{1}{k}-k\right) =\hat{\Delta}'_{inner } = \quad \quad \quad \quad \quad\\
         = \beta N_0  \sqrt{ \left(\frac{\gamma_{\mathrm{lin}}}{A_0 k}\right)}   \left[ \frac{2.12}{\sqrt{\beta}}   +  \pi  \bigg ( \frac{\frac{1}{\varphi}+\mu_e N_0^2}{-\mu_e N_0^2} \bigg ) \bigg ( \frac{\gamma}{\lambda} \frac{-\mu_e N_0^2}{A_0 k} \bigg )^{1/2} \right]    
         \end{split}
\end{align}

\bibliography{aipsamp}% Produces the bibliography via BibTeX.

\end{document}